\documentclass{aa}  

\usepackage{graphicx}
\usepackage{placeins}
\usepackage{ulem}
\usepackage{xcolor}
\usepackage{array}
\usepackage{txfonts}
\usepackage{subfig}

\usepackage{natbib}

\usepackage{hyperref}[links=true, link={red}, cite={blue}]
\hypersetup{
    colorlinks = true,
    linkcolor = {red},
    citecolor = {blue}
}
\begin{document} 

\let\textbf\relax

   \title{Asteroid reflectance spectra from $Gaia$ DR3: Near-UV in primitive asteroids\thanks{Tables with the computed parameters of the 0.7 $\mu$m band are only available in electronic form at the CDS via anonymous ftp to \url{cdsarc.u-strasbg.fr} (130.79.128.5) or via \url{http://cdsweb.u-strasbg.fr/cgi-bin/qcat?J/A+A/}}}

   \author{F. Tinaut-Ruano
          \inst{1,2}
          \and
          J. de Le\'{o}n\inst{1,2}
          \and
          E. Tatsumi\inst{3}
          \and
          D. Morate\inst{4}
          \and
          M. Mahlke\inst{5}
          \and
          P. Tanga\inst{6}
          \and
          J. Licandro\inst{1,2}
          }

   \institute{Instituto de Astrofísica de 
     Canarias (IAC), C/ Vía Láctea, s/n, E-38205, La Laguna, Spain\\
     \email{fernando.tinaut@iac.es}
     \and
     Department of Astrophysics, University of La Laguna, Tenerife, Spain
     \and 
     Institute of Space and Astronautical Science (ISAS), Japan Aerospace Exploration Agency (JAXA), Sagamihara, Kanagawa, Japan
     \and
     Centro de Estudios de Física del Cosmos de Aragón (CEFCA), Plaza San Juan, 1, E-44001 Teruel, Spain
     \and 
     Institut d'Astrophysique Spatiale, Université Paris-Saclay, CNRS, F-91405 Orsay, France.
     \and
     Universit\'{e} C\^{o}te d'Azur, Observatoire de la C\^{o}te d'Azur, CNRS, Laboratoire Lagrange, Bd de l'Observatoire, CS 34229, 06304 Nice Cedex 4, France
     }

   \date{Received 27/11/2023; accepted 27/02/2024}

 
  \abstract
   {In the context of charge-coupled devices (CCDs), the ultraviolet (UV) region has mostly remained unexplored after the 1990s. Gaia DR3 offers the community a unique opportunity to explore tens of thousands of asteroids in the near-UV as a proxy of the UV absorption. This absorption has been proposed in previous works as a diagnostic of hydration, organics, and space weathering.}
   {In this work, we aim to explore the potential of the NUV as a diagnostic region for primitive asteroids using Gaia DR3.}
   {We used a corrective factor over the blue part of $Gaia$ spectra to erase the solar analog selection effect. We identified an artificial relation between the band noise and slope and applied a signal-to-noise ratio (S/N) threshold for $Gaia$ bands. Meeting the quality standards, we employed a Markov chain Monte Carlo (MCMC) algorithm to compute the albedo threshold, maximizing primitive asteroid inclusion. Utilizing one- and two-dimensional (1D and 2D) projections, along with dimensionality-reduction methods (such as PCA and UMAP), we identified primitive asteroid populations.}
   {We uncovered: (a) the first observational evidence linking UV absorption to the 0.7 $\mu$m band, tied to hydrated iron-rich phyllosilicates; and (b) a 2D space revealing a split in C-type asteroids based on spectral features, including UV absorption. The computed average depth (3.5 $\pm$ 1.0 \%) and center (0.70 $\pm$ 0.03 $\mu$m) of the 0.7 $\mu$m absorption band for primitive asteroids observed with $Gaia$ is in agreement with the literature values.}
   {In this paper, we shed light on the importance of the UV absorption feature to discriminate among different mineralogies (i.e., iron-rich phyllosilicates vs. iron-poor) or to identify taxonomies that are conflated in the visible (i.e., F-types vs. B-types). We have shown that this is a promising region for diagnostic studies of the composition of primitive asteroids.} 


   \maketitle
%

\section{Introduction}\label{sec:int}

The study of primitive asteroids is essential to understanding the formation and evolution of the Solar System. These asteroids are believed to be remnants of the early stages of planetary formation, containing the most pristine material of the Solar System. Thus, studying the composition of  primitive asteroids is the closest approach to the original composition of the protoplanetary disk.

In the visible to the near-infrared (up to 2.5 $\mu$m), the spectra of primitive asteroids are typically featureless, with the sole exception of a shallow absorption band centered at 0.7 $\mu$m, associated with the presence of Fe-rich phyllosilicates \citep{1994Icar..111..456V}. The most diagnostic wavelength region for primitive asteroids is near 3-$\mu$m, where absorption bands, due to hydroxyl OH and centered around 2.7-2.9 $\mu$m, \citep{2019GeoRL..4614307H} or even water ice, centered at $\sim$3.1-3.2 $\mu$m, are present \citep{2015aste.book...65R}. Intriguingly, another characteristic of primitive asteroids is their spectral behavior at wavelengths below $\sim$0.5$\mu$m, where a drop-off in the reflectance could appear. It is caused by an absorption band in the ultraviolet (UV) due to a ferric oxide intervalence charge transfer (IVCT)  and associated with Fe-bearing phyllosilicates \citep{1989Sci...246..790V}. However, due to Earth's atmosphere, it is impossible to observe both regions completely from the ground and, thus, space-based observations are needed to properly study them. 

However, we have recently shown a tentative but promising connection between the NUV absorption and the 0.7 $\mu$m and 3 $\mu$m absorptions associated with Fe-rich phyllosilicates, indicating that the NUV region might be used as a proxy for hydration in primitive asteroids using ground-based data from SDSS and ECAS surveys, as well as NUV to visible spectra obtained with the 3.6-m Telescopio Nazionale Galileo (TNG), located on the island of La Palma, Spain \citep{2022A&A...664A.107T,2023A&A...672A.189T}. Those previous works motivated the present study, where we aim to further explore this connection by investigating the spectral behavior of primitive asteroids observed from the  $Gaia$ space telescope  in the NUV. The paper is organized as follows. In Section \ref{sec:sam}, we describe the $Gaia$ dataset and the selection criteria for our samples. Section \ref{sec:met} presents the methodology used to analyze the samples, including the definition of several features that we measured over the reflectance spectra (slopes and band parameters) and the unsupervised dimensionality-reduction algorithms that we used. In Section \ref{sec:res} we present and discuss the obtained results. Finally, in Section \ref{sec:con}, we summarize our findings and describe their implications for future works.

\section{Dataset}\label{sec:sam}
We used the reflectance spectra of asteroids provided by $Gaia$ DR3 \citep{2023A&A...674A..35G}. The  reflectance spectra were obtained by dividing each epoch spectrum by the mean of the solar analog stars selected and then averaging over the set of epochs. The resulting spectrum consists of 16 reflectance values (hereafter, also referred to as "bands") covering a wavelength range from 0.374 to 1.034 $\mu$m, with a resolution of 0.044 $\mu$m, and normalized to unity at $0.550 \pm 0.025$ $\mu$m.
From the original number of objects provided by $Gaia$ DR3 (60,518), more than a third (26,577) have data for all 16 bands. However, less than a sixth (9,926) have all the bands observed with the best quality flag for every one of the bands.

As we are looking for as much information as available for primitive asteroids among the 60,518 asteroids in $Gaia$, our first selection criterion is to use the albedo to identify primitive ones. We obtained the albedo value, available for 41,720 asteroids, from the SsODNet \citep{2023A&A...671A.151B} using the Python tool $Rocks$\footnote{\url{https://github.com/maxmahlke/rocks}}. Various authors have used slightly different values for the albedo to distinguish among primitive and non-primitive asteroids, with the typical value being around 10\% \citep{1984PhDT.........3T, 2010ApJ...721L..53C, 2023A&A...672A.189T}. We searched for the best albedo value to select the primitives among the $Gaia$ sample.

\begin{figure}
    \centering
    \includegraphics[width=0.49\textwidth]{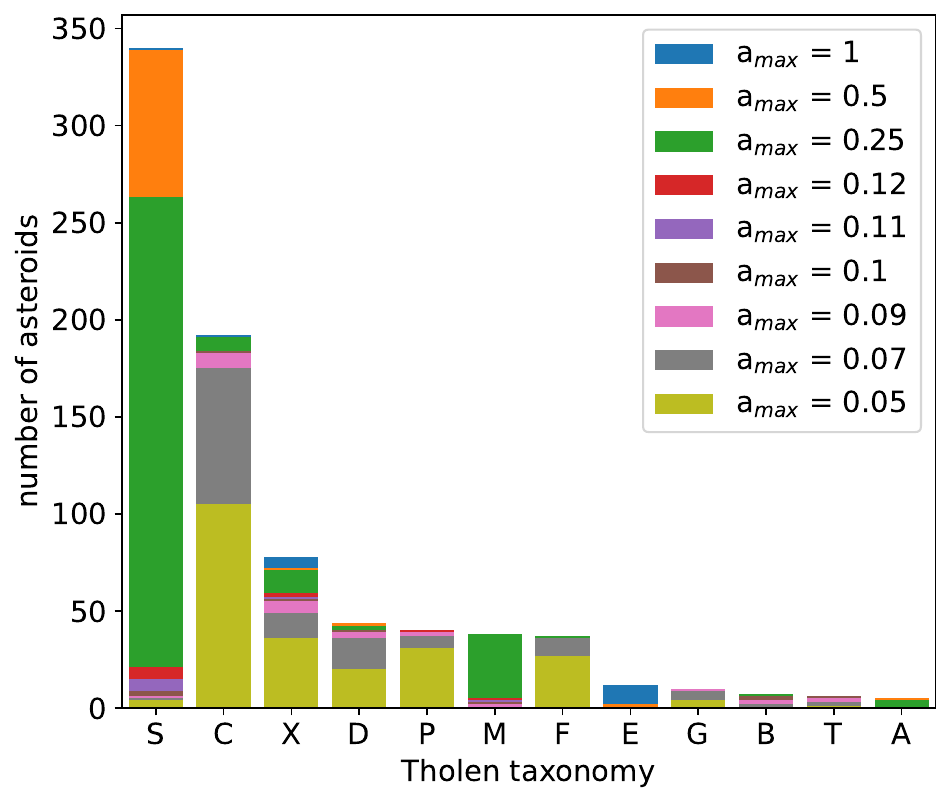}
    \caption{Remaining objects classified in \cite{1984PhDT.........3T} with Gaia spectra after different albedo thresholds are applied (a$_{max}$). The threshold a$_{max}$=0.105 maximizes the number of primitives.}
    \label{fig:alb}
\end{figure}

In Fig. \ref{fig:alb}, we show the number of asteroids that were classified in \cite{1984PhDT.........3T}, with available $Gaia$ spectra, which have an albedo value lower than a given albedo threshold (a$_{max}$). All the taxonomies that we refer to hereafter are based on \cite{1984PhDT.........3T}. By taking a$_{max}$=0.5 we can separate mainly E-types. With  a$_{max}$=0.12, we lose 93\% of the S-, M-, A-, and V-types. Only when we go down to a$_{max}$=0.07, we lose 11\% of the original sample of C, P, D, F, B, G, and T asteroids, which are what we consider primitive in this work. Applying a Markov chain Monte Carlo approach, in which we randomly selected 80\% of the total sample, over 1000 iterations, and computed the albedo which maximizes the number of primitive asteroids over non-primitives (S, M, E, A, and V), we obtain an albedo threshold of 0.105 $\pm$ 0.017. From here on, we only deal with the 15,529 asteroids with known albedo smaller than 0.105 and for which $Gaia$ measured a spectrum. Literature sources used to compile both albedo and diameter for those asteroids are listed in Table \ref{sec:app_ref}. X-types were not used for the albedo calculation as they include E-, M-, and P-types having no albedo information at the time Tholen classified them. P-types were defined in the Tholen taxonomy as low albedo X-types. Thus, the remaining X-types present in the sample after we applied the albedo cut-off, were re-classified as P-types.

\begin{figure}
    \centering
    \includegraphics[width=0.49\textwidth]{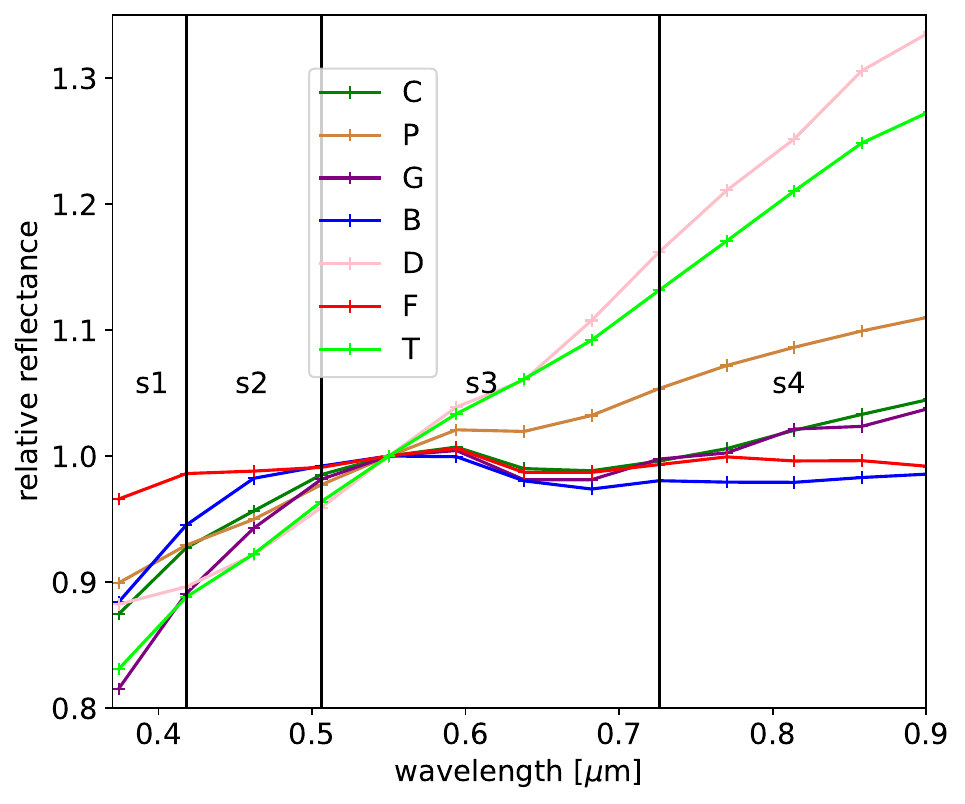}
    \caption{Median of $Gaia$ spectra for each primitive taxonomy, with the wavelength regions used to compute each spectra slope marked as black vertical lines.}
    \label{fig:ref_sp}
\end{figure}

Once we have separated primitive asteroids from the total sample, we computed several spectral slopes by least square fitting of the normalized reflectance in different wavelength ranges (see Fig. \ref{fig:ref_sp}): from 0.374 to 0.418 $\mu$m ($s1$); from 0.418 to 0.506 $\mu$m ($s2$); from 0.506 to 0.726 $\mu$m ($s3$); and from 0.726 to 0.902 $\mu$m ($s4$). If a band is missing or the quality flag is not the best in the particular wavelength range of a spectral slope, we decided not to compute it. That leaves 3,778 primitive asteroids with $s1$, 14,582 with $s2$, 15,084 with $s3$, and, finally, 15,170 have $s4$ computed.

\begin{figure}
    \centering
    \includegraphics[width=0.49\textwidth]{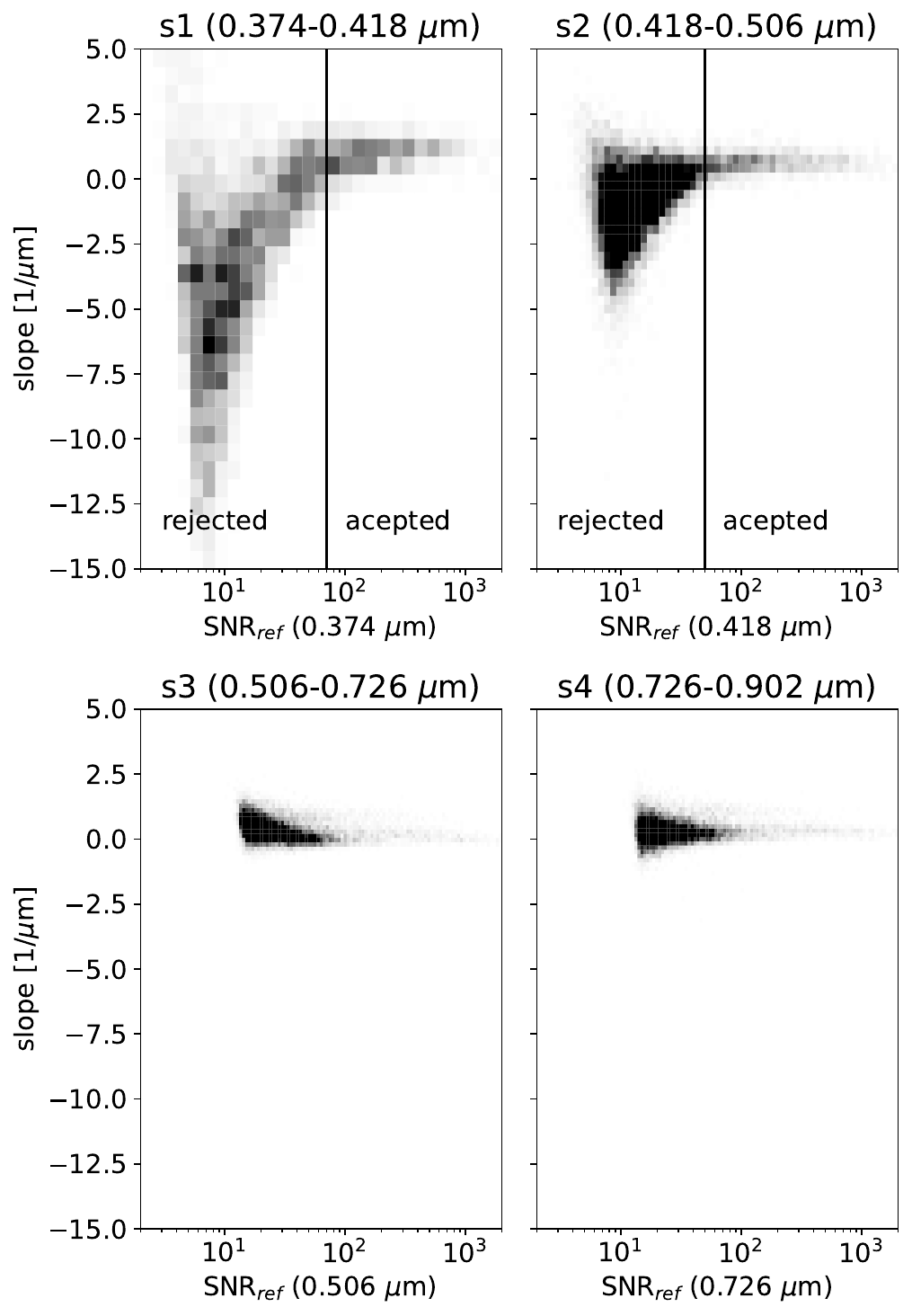}
    \caption{Density plots of the computed slopes versus $\rm{S/N_{ref}}$ at different bands. Vertical black lines indicate the minimum S/N$_{\rm ref}(\lambda)$ to be selected.}
    \label{fig:snr}
\end{figure}

In the process of analyzing the quality of the data, we computed the parameter $\rm{S/N_{\rm ref}}(\lambda) = R(\lambda)/R_{\rm error}(\lambda)$, where $R(\lambda)$ is the normalized reflectance value at one particular band, and $R_{\rm error}(\lambda)$ is its associated error calculated by \citep{2023A&A...674A..35G}. When searching for potential systematics, we discovered a trend in $s1$ and $s2$ versus $\rm{S/N_{ref}}$, shown in the upper panels of Fig. \ref{fig:snr}. We observe a decrease in those slopes with decreasing values of $\rm{S/N_{ref}}$, which seems to be an artifact and unrelated to any other physical parameter (albedo, diameter, orbital parameters, or taxonomy). On the contrary, the values of $s3$ and $s4$ do not present any tendency (see lower panels of Fig. \ref{fig:snr}), and their variation, ranging from -1 to 2 $\mu$m$^{-1}$, is consistent to the ones observed in other spectroscopic surveys, as in \cite{2002Icar..158..106B}. As we are not able to explain nor correct the trends observed for $s1$ and $s2$, we decided to use the $\rm{S/N_{ref}}$ value as a threshold: the selected value is such that we minimize the number of asteroids showing the trend and we maximize the number of asteroids with reliable values of the spectral slopes. Therefore, we calculated $s1$ only for those asteroids with $\rm{S/N_{ref}}(0.374) > 70$, resulting in a total of 522 asteroids. In the same way, $s2$ is calculated for the 1,331 asteroids, which have a signal-to-noise ratio of S/N$_{ref}$(0.418) $>$ 50.

Apart from this observed trend, the last three bands after 0.902 $\mu$m are often affected by large random errors \cite{2023A&A...674A..35G}. As we will not compute anything beyond 0.902 $\mu$m, we did not use those bands. Furthermore, $Gaia$ spectra came from two different photometers (red and blue) and the band at 0.632 $\mu$m is problematic for bright objects (see Ceres and Vesta spectra at Fig. 13 and 14 of \cite{2023A&A...674A..35G}) as it is the joining region between them \citep{2023A&A...674A..35G}. Despite this problem, we decided to use this particular band in our computations as it is close to the center of the 0.7 $\mu$m absorption band, a key feature in our study. We tested all our computations with and without this band. When we used it, as it has more variance because of its systematics, the loading in the principal component base  is overestimated for this band (see Fig \ref{fig:PCA}) but the principal component space is still similar. Nevertheless, without the band at 0.632 $\mu$m, many more fake detections of the 0.7 $\mu$m band are present in our algorithms.

\section{Methodology}\label{sec:met}

Before starting any computation, we corrected the four bluest bands of $Gaia$ spectra. The selection of solar analogs by the $Gaia$ team included a large number of the most commonly used stars by the planetary science community \citep{2023A&A...674A..35G}. Such stars can be used to get asteroid reflectance spectra at wavelengths beyond 0.5 $\mu$m without any problems. However, as we have shown in \cite{2022A&A...664A.107T}, they do not have a Sun-like spectral behavior at the wavelengths below 0.5 $\mu$m, introducing an artificial reddening in that region. As we are interested in the near-UV part of the $Gaia$ spectra, we computed a multiplicative correction factor in \cite{2023A&A...669L..14T}, which we applied to all asteroid spectra used in this work.

To study the properties of primitive asteroids we used all the available information that we can extract from the $Gaia$ spectra, as well as other physical parameters available. We use the spectral slopes defined in the previous section and shown in Fig. \ref{fig:ref_sp}, as well as combinations having physical meaning. For example, the value ($s1-s2$) indicates how much the slope changes around 0.418 $\mu$m, as it does ($s2-s3$) for 0.506 $\mu$m. Both of them are indeed describing the amount of drop-off in NUV reflectance, namely, the beginning of the UV absorption. As another example, the value ($s4-s3$) is related to the curvature of the spectra around $\sim$0.7 $\mu$m, namely, the presence of the 0.7 $\mu$m absorption band. 

We searched for other available information on the asteroids. Albedo values, previous taxonomical classification, and diameters were obtained using the Python tool $Rocks$ from \cite{2023A&A...671A.151B}. To check if an asteroid belonged to a particular collisional family, we used the lists from \cite{2015aste.book..297N}. To look for trends, clusters, and/or correlations we used different techniques. We applied principal component analysis (PCA) and uniform manifold approximation and projection (UMAP) to the asteroid reflectance spectra, but also to a set of spectral parameters: $s$, defined as the spectral slope over the NUV-visible wavelength range (0.418 to 0.902 $\mu$m), $s2$, ($s2$-$s3$), and ($s4$-$s3$). The choice of these particular parameters will be explained in Section \ref{sec:res}.

The PCA is a dimension reduction technique that aims to transform a dataset into a new coordinate system by finding orthogonal linear combinations of the original features, called principal components, that retain as much information as possible while reducing the dimensions. The variance of the data along each axis is maximized, with the first principal component capturing the most variance, the second one capturing the second most, and so on. Previous authors have used PCA to define and visualize the distribution of taxonomic classes and find the combination of reflectances that describes the spectral features \citep{1984PhDT.........3T, 2002Icar..158..106B}. We carried out two different PCAs: the first one tried to replicate the one done by \cite{2002Icar..158..106B} but used only primitive asteroids. Thus, we flattened the spectra by dividing it by the slope computed as a least squared fit of the reflectances between 0.418 $\mu$m to 0.902 $\mu$m and used those slope-removed reflectances without standard deviation nor mean scaling for the PCA (following Bus\&Binzel's notation, we refer to these components as PC'). The second one does not use reflectances, but four parameters that describe the spectral shapes of the different asteroid taxonomies. These parameters are: visible slope $s$, as computed in \cite{2002Icar..158..106B}, blue slope $s2$, change in slope at 0.5 $\mu$m ($s2$-$s3$), and change in slope at 0.7 $\mu$m ($s4$-$s3$) (see Fig \ref{fig:ref_sp}). As these parameters have different value ranges, we have to scale the dataset before doing the PCA. The standard scaling of each set of variables changes their distribution closer to a normal distribution by subtracting each mean and dividing by each standard deviation.  Both PCAs are applied over 1,331 asteroids as we are discarding the bluest filter for this study.

The uniform manifold approximation and projection (UMAP) is another dimensionality-reduction algorithm, using non-linear feature coefficients. It tries to learn the manifold structure of the data and find low-dimensional representations (usually 2D) that preserve the topological structures of the original data. That means that points that are closer in the 2D representations are also closer in the $n$-dimensional space, with n=4 in this case ($s$, $s2$, $s2$-$s3$, and $s4$-$s3$). The combination of those 4 variables, to arrive at the 2D space, does not have to be linear. Further information on UMAP can be found in \cite{sainburg2021parametric}. To implement it we used the library \textit{umap}\footnote{\url{https://umap-learn.readthedocs.io/en/latest/}} for Python with the parameters: \textit{n\_neighbors}=100, \textit{min\_dist}=0, and \textit{random\_state}=0.

To compute the band depth for the 0.7 $\mu$m absorption band, we fit a model based on a flat spectrum normalized at 0.55 $\mu$m and having a Gaussian absorption in the wavelength range from 0.55 to 0.91 $\mu$m. Usually, these parameters were computed by subtracting a continuum and then fitting a Gaussian or a polynomial to the absorption band. However, due to the low resolution of the $Gaia$, in this classical approach, the continuum is fitted using only two or three points in the shoulders of the band, and this introduces errors and unrealistic band depth measurements for a large fraction of the asteroid sample. Our model uses instead all the available $Gaia$ bands in the wavelength range and thus, reduces the errors and provides more reliable measurements. The chosen model can be described as:
\begin{equation}\label{eq:fitting_model}
    r(\lambda)=(m\times\lambda+1-m\times 0.55)\left(1-de^\frac{-(\lambda - \lambda_c)^2}{2w^2}\right)
,\end{equation}
where $r(\lambda)$ is the normalized reflectance, $m$ is the slope of the continuum fitted across the entire wavelength range (in $\mu$m$^{-1}$), $d$ is the band depth (in values from 0 to 1), $\lambda_c$ is the center of the 0.7-$\mu$m band (in $\mu$m), and $w$ is the width of the band computed as the standard deviation of the Gaussian. The free parameters on this model are $m$, $d$, $\lambda_c$, and $w$. To avoid overfitting of Gaussians between Gaia spectral bands we decided to establish a lower limit of 0.035 $\mu$m in $w$, a value that is 50\% larger than half of the wavelength resolution (i.e., 0.022 $\mu$m). The upper limit is based on a visual inspection of tens of $Gaia$ spectra, where we see that the wider absorption bands (that are not simply concavity of the spectra) have a  $w$ value of 0.085 $\mu$m. 

\section{Results and discussion}\label{sec:res}
In this section, we present the distribution of the computed spectral slopes, as well as their combinations, for the final sample of 15,529 primitive asteroids in $Gaia$ DR3 with an albedo in the literature (references at \ref{tab:alb_ref}). Density plots relating different parameters can also help to obtain relations between variables or to identify differences between groups of asteroids. Objects already classified are also helpful in identifying features in density plots and associating them with taxonomies. We present also the results of applying both PCA and UMAP algorithms, and any correlation with taxonomy, diameter, albedo, and membership to collisional families. Taxonomic classifications were obtained from \cite{1984PhDT.........3T} as it is the work with the most similar wavelength coverage and spectral resolution to $Gaia$. Asteroids having uncertain or multiple taxonomic classifications are not considered here. 

\subsection{General behaviors of primitive asteroids}\label{ssec:res_gen}
\begin{figure*}
    \centering
    \includegraphics[width=0.9\textwidth]{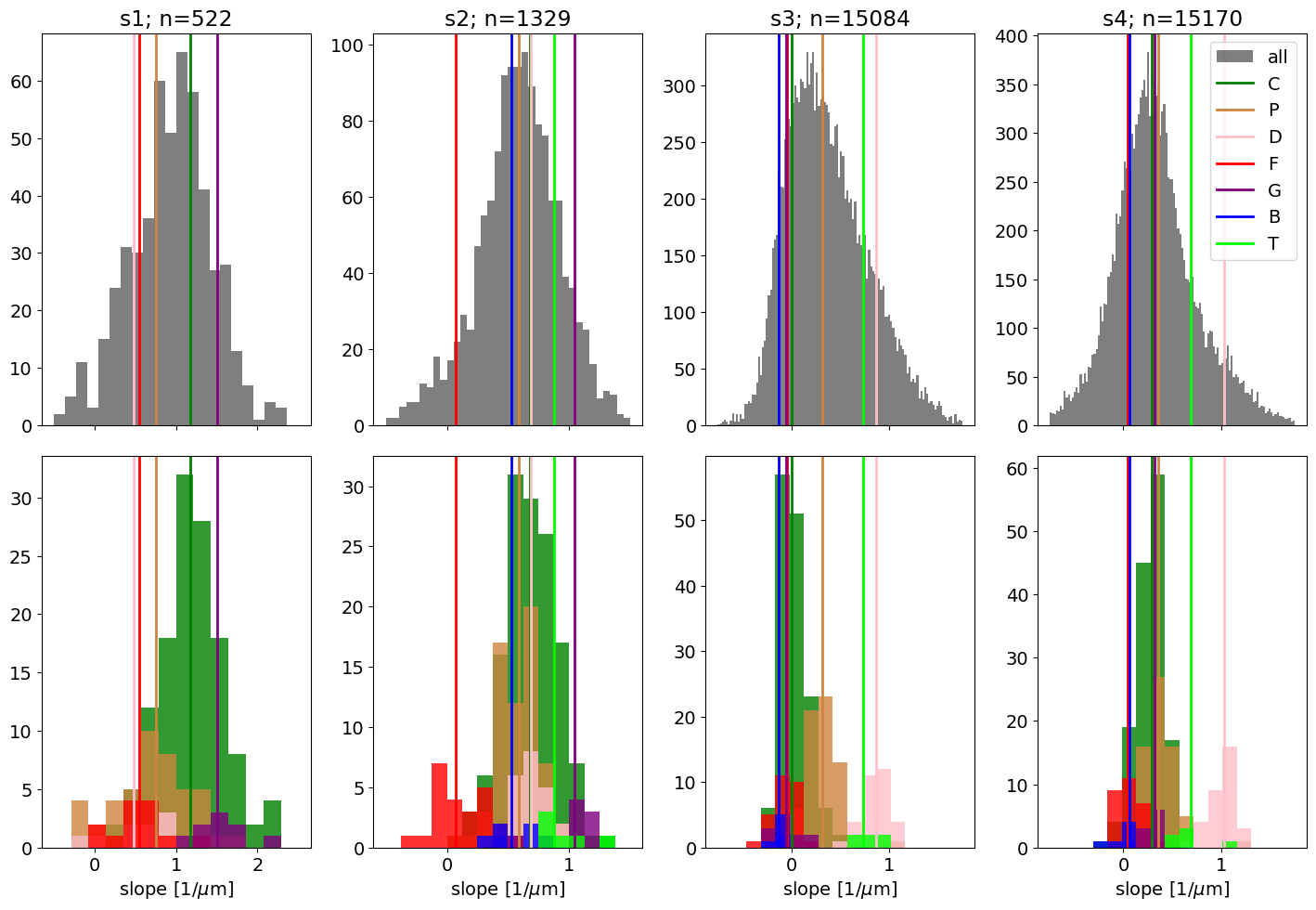}
    \caption{Slope distributions (from $s1$ at left to $s4$ at right) for all the asteroids in our sample (upper panel) and for those classified by \cite{1984PhDT.........3T} (lower panel). Vertical lines indicate the computed median for each taxonomical class. The title of each column indicates the total number of asteroids with such slope computed. B- and T-types have five or fewer members with computed $s1$, thus, we decided not to show them in the corresponding plots.}
    \label{fig:slope_hist}
\end{figure*}

We present the distribution of computed slopes for our sample of asteroids in Fig. \ref{fig:slope_hist}. The upper panel shows the slope distributions for all the primitive asteroids in our sample (in gray), and the lower panel shows the same but for those asteroids classified by Tholen as C, P, D, F, B, T, and G. Vertical lines indicate the median values of each taxonomic class computed using the $Gaia$ data and shown in Table \ref{tab:slopes}, together with the number of asteroids for each taxonomy in between parentheses. The number of bins for the histograms is computed as the integer part of the square root of the number of asteroids available for each slope.

\begin{table*}
\caption{Median values of slopes for each taxonomy.}
\label{tab:slopes}
\centering
\begin{tabular}{c c c c c c c c}   
\hline\hline\\[-3mm]
Slope & \multicolumn{7}{c}{Taxonomy }\\
\hline\\[-3mm]
[1/$\mu$m] & C (\#)     & G (\#)        & P (\#)    & T (\#)        & D (\#)        & B (\#)        & F (\#)  \\
\hline\\[-3mm]
$s1$    & 1.18 (130)    & 1.50 (9)      & 0.76 (45) & -- (5)        & 0.48 (11)     & -- (4)        & 0.55 (13)\\
$s2$    & 0.68 (138)    & 1.04 (10)     & 0.59 (64) & 0.88 (6)      & 0.68 (22)     & 0.52 (6)      & 0.07 (24)\\
$s3$    & 0.01 (145)    & $-0.05$ (9)   & 0.32 (67) & 0.74 (6)      & 0.87 (32)     & $-0.13$ (6)   & $-0.04$ (28)\\
$s4$    & 0.29 (145)    & 0.32 (10)     & 0.35 (67) & 0.69 (6)      & 1.03 (33)     & 0.06 (6)      & 0.04 (28)\\
($s1-s2$) & 0.50 (130)  & 0.47 (9)      & 0.12 (45) & -- (5)        & $-0.22$ (11)  & -- (4)        & 0.39 (13)\\
($s2-s3$) & 0.67 (138)  & 1.01 (9)      & 0.25 (64) & 0.18 (6)      & $-0.15$ (22)  & 0.61 (6)      & 0.14 (24)\\
($s4-s3$) & 0.23 (145)  & 0.37 (9)      & 0.07 (67) & $-0.20$ (6)   & 0.11 (32)     & 0.15 (6)      & 0.10 (28)\\
\hline
\end{tabular}\\
\footnotesize{\#: In between parentheses are the number of asteroids used to compute the median.\\--: The median is not computed for those slopes where 5 or fewer asteroids were available.}
\end{table*}

The spectral slope space is useful for separating some of the taxonomies and inspecting their general behavior. D-types (in pink) have slightly red spectra in the NUV (low $s1$), getting redder as moving towards longer wavelengths, with a very red slope (very high values of $s3$ and $s4$) that makes them quite easy to differentiate from the rest of the classes. F-types (in red) are flat in the whole spectral range, but the $s1$ shows a slight red slope (i.e., the positive slope at the wavelengths below 0.418 $\mu$m). This could be interpreted as the F-types having the beginning of the UV absorption around 0.418 $\mu$m. They also show the flattest spectral slopes in $s2$, where they separate from the other taxonomies. For $s3$ and $s4$, P, T, and D types, have the largest slope values (in this order) of all the primitive taxonomies. On the other hand, G-types have the largest $s2$, namely, the UV absorption starts below 0.506 $\mu$m. The largest values of $s1$ are for G-types as they have the deepest UV absorptions. C-types have a similar behavior to G-types but with less pronounced UV absorption. P-types have a slightly decreasing slope with increasing wavelength. B-types have a similar behavior to F-types, but the transition from red slope to flat is observed at 0.506 $\mu$m, indicating that the beginning of the UV absorption is at longer wavelengths than that of F-types.

\begin{figure*}
    \centering
    \includegraphics[width=\textwidth]{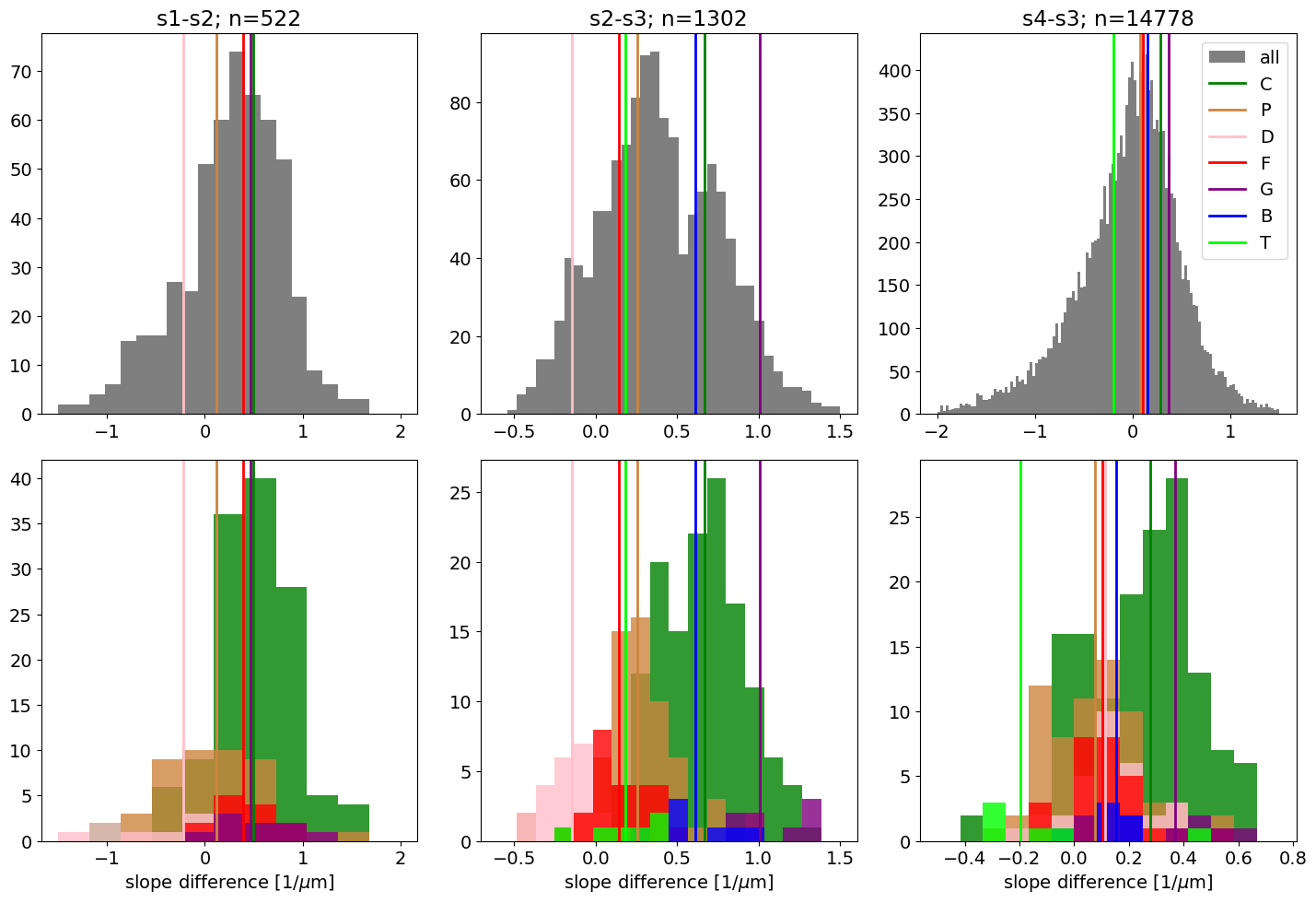}
    \caption{Distribution of slopes for the primitive asteroids in our sample (upper panel) and those classified by \cite{1984PhDT.........3T} (bottom). Vertical lines indicate the computed median for each taxonomical class. The title of each column indicates the total number of asteroids with such difference computed. B- and T-types have less than five members with computed $s1$, thus, we decided not to show them in the ($s1$-$s2$) plot. The rightest column does have not a shared x-axis as the general ($s3$-$s4$) distribution is much wider than the classified ones.}
    \label{fig:slope-slope_hist}
\end{figure*}


Combining some of the computed slopes allows us to better separate between those taxonomies that have a change in slope in a certain wavelength (see Fig. \ref{fig:slope-slope_hist}). For example, all the G-types and some C-types have an absorption band at 0.7 $\mu$m, which translates into their high values of $s4$-$s3$ (right column of Fig. \ref{fig:slope-slope_hist}). The negative median value of ($s2-s3$) for the D-types indicates a spectral flattening below 0.5 $\mu$m, in agreement with \cite{2003Icar..164..104E, 2019AJ....157..161W}, which is a unique and reveals the opposite behavior with respect to the rest of primitive taxonomies. F-types show a larger value for ($s1-s2$) than for the other combinations, as is expected from their larger values of $s1$ and the close value to 0 for the rest of the slopes. That shows that their UV absorption begins at 0.4 $\mu$m. Finally, ($s2-s3$) shows a bi-modality in the general distribution (top center panel of Fig. \ref{fig:slope-slope_hist}). Large values of ($s2-s3$) seem to be related to a change in spectral slope around 0.5 $\mu$m for the B-types, and the presence of a 0.7 $\mu$m absorption band together with strong UV absorption for the G-types and some C-types.

\begin{figure}
    \centering
    \includegraphics[width=0.49\textwidth]{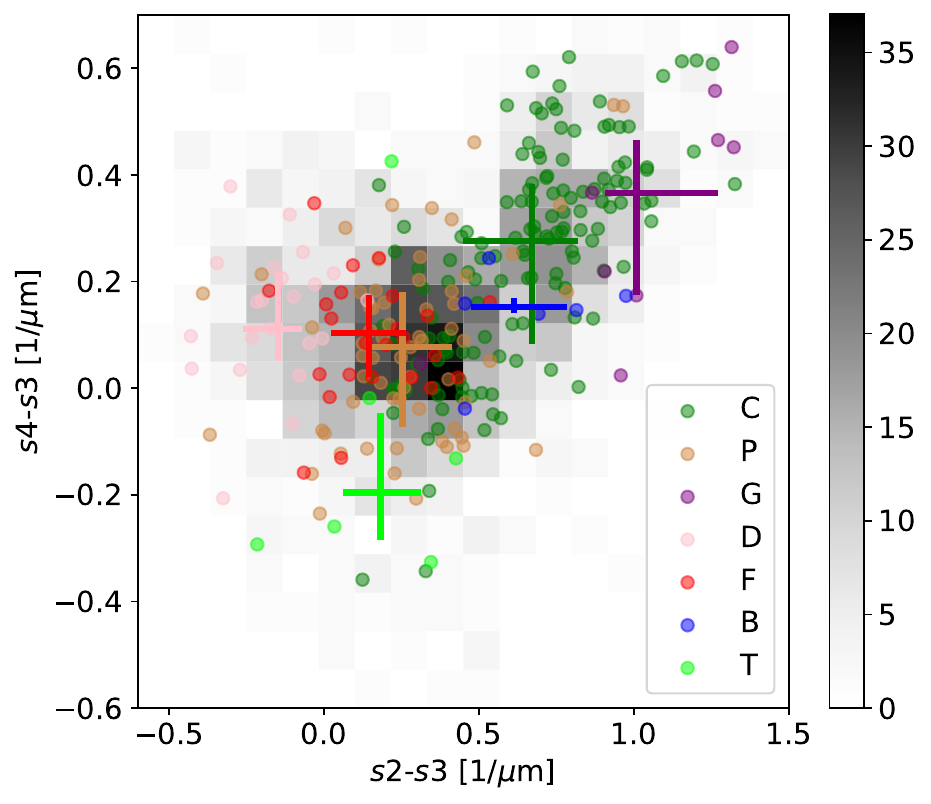} \includegraphics[width=0.49\textwidth]{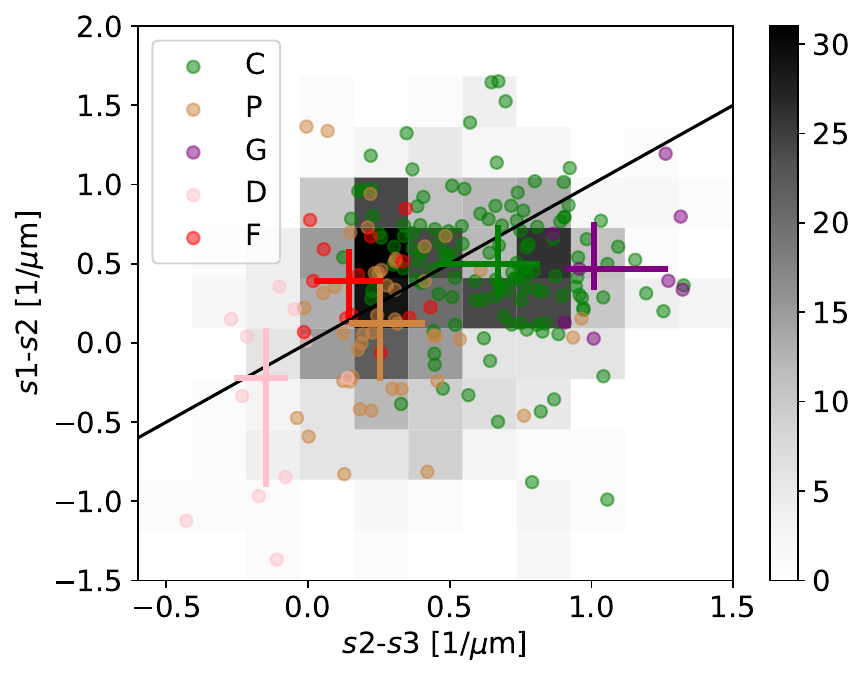}
    \caption{Density plot of ($s2-s3$) vs. ($s4-s3$) for the 1,241 asteroids with these computed slopes in $Gaia$ DR3 (upper panel). Colors represent asteroids classified by \citet{1984PhDT.........3T}. The crosses mark the median of each taxonomy and the size of the error bars is the interquartil range. The lower panel shows the same density plot, but for ($s2-s3$) vs. ($s1-s2$). In this case, a total of 498 asteroids had those slopes computed in $Gaia$ DR3. We only include here the taxonomies with more than five members with computed $s1$ (i.e. C, P, G, D, and F). }
    \label{fig:corr}
\end{figure}

In Figure \ref{fig:corr}, we plot ($s2-s3$) vs. ($s4-s3$) in the upper panel and ($s1-s2$) vs. ($s2-s3$) in the lower panel using density plots (gray hatches) of all available asteroids with those differences computed. Asteroids classified by \citet{1984PhDT.........3T} are marked with colored dots representing different taxonomies, the crosses indicate the median value of each taxonomy using the same color code, and the size of the error bar is the interquartile range. In the upper panel, we can see the correlation between ($s2-s3$), which is related to the absorption in the UV, and ($s4-s3$), which can be considered as a proxy of the 0.7-$\mu$m band. We note that for the case of D and F types, which do not present usually the 0.7-$\mu$m feature, they have small but positive values of ($s4-s3$). Those values (and a visual inspection) suggest a straight to slightly concave-up-shaped spectrum on both taxonomies. The existence of concave-up and concave-down D-types has been previously noticed by different authors  \citep{1987Icar...72..304B, 1994A&A...282..634F, 2003Icar..161..356C}. Meanwhile, the F types are generally negative in $s3$ and flat in $s4$, generating this slightly concave-up shape.

We searched for potential correlations between ($s2-s3$) and ($s4-s3$) using a simple Pearson correlation coefficient (PCC). We found a PCC value of 0.7 for the asteroids with P, C, and G taxonomies. Apart from that, F and D taxonomies have the smallest values of ($s2-s3$), even with negative values (i.e., smaller UV absorption up to 0.4 $\mu$m). That tells us again that their spectra are getting bluer towards short wavelengths, at least until 0.4 $\mu$m. On the other hand, G-types have the strongest UV absorption (larger $s2-s3$ values) and the deeper 0.7-$\mu$m band (larger $s4-s3$ values). That is expected as G types were described by Tholen as primitive asteroids with large UV absorption and they usually present the 0.7 $\mu$m \cite{1984PhDT.........3T, 2002Icar..158..106B}. P-types distribute around the central part of the density plot, between C-G and F-D taxonomies. They present a red to flat spectra with increasing wavelength. One part of the bi-modal C-type distribution overlaps with the non-hydrated P-types, while the other part overlaps with the hydrated G-types. This separation is decreased without the NUV part of the spectra as shown in \cite{2022A&A...665A..26M} for other equivalent taxonomies. C-types present the hydration band and a stronger UV absorption more often than P types, but with shallower bands than G types, as defined by \cite{1984PhDT.........3T}.

Inspecting now the general density distribution we see a maximum in density populated by P-, F-, and, some C-types. Following the correlation, the most populated area to the upper right is mainly populated by C- and G-types. This high population seems to be related to the bi-modality in the ($s2-s3$) seen in Fig. \ref{fig:slope-slope_hist}. From the central maxima to the left, we found a slight overdensity presumably related to D- and F-types.

In the lower panel of Fig. \ref{fig:corr}, we explore the UV absorption more in detail. The unity line (black line) where objects that have the same change in slope at 0.5 $\mu$m as at 0.4 $\mu$m would fall, namely, a constant absorption, including linear spectral shape located around (0, 0). C-, P- and D-types are on the unity line, with C- types having a huge dispersion in the space but with positive values in both slope differences with slightly smaller ($s1-s2$; seen as a concave-down shape between 0.374 and 0.726 $\mu$m); P-types also have positive values in both slope differences but smaller than C-types; D-types have negative values for both slope differences, which translates into a concave-up shape in the same wavelength range, i.e., a flattening towards bluer wavelengths. The only taxonomy with a median value above the unity line is the F-type, showing that the change in slope is higher at 0.4 $\mu$m than it is at 0.5 $\mu$m. Below the unity line, we found G-types with larger values of ($s2-s3$) than ($s1-s2$), which implies a stronger increase in slope for decreasing wavelength around 0.5 $\mu$m than around 0.4 $\mu$m. Taking a look at the general distribution, we can see two local maxima, one below the unity line close to (0.75, 0.5), and another above the line, close to the F-type median. This bimodality suggests the existence of two different groups: one with larger values of ($s2-s3$) and slight dispersion on ($s1-s2$), indicating that the beginning of the UV absorption is at 0.5 $\mu$m and the decrease in reflectance continues towards 0.4 $\mu$m (G-types and some C-types); the other with lower values of ($s2-s3$) and with a higher dispersion on ($s1-s2$), ranging from small or negative values (i.e. P- and D-types, respectively) to large values of ($s1-s2$), which indicate a UV absorption starting at 0.4 $\mu$m (F-types). \textbf{\cite{1978SSRv...21..555G} have suggested that the location of the UV absorption edge is diagnostic of composition for meteorites with albedo matching C-type asteroids, an absorption beginning at 0.5 $\mu$m reveal the presence of iron-rich clay mineral, while an absorption edge at 0.4 $\mu$m could be due to either an iron-poor clay mineral or iron-rich olivine and pyroxene.}

\subsection{Dimensional reduction: PCA and UMAP}
\begin{figure}
    \centering
    \includegraphics[width=0.49\textwidth]{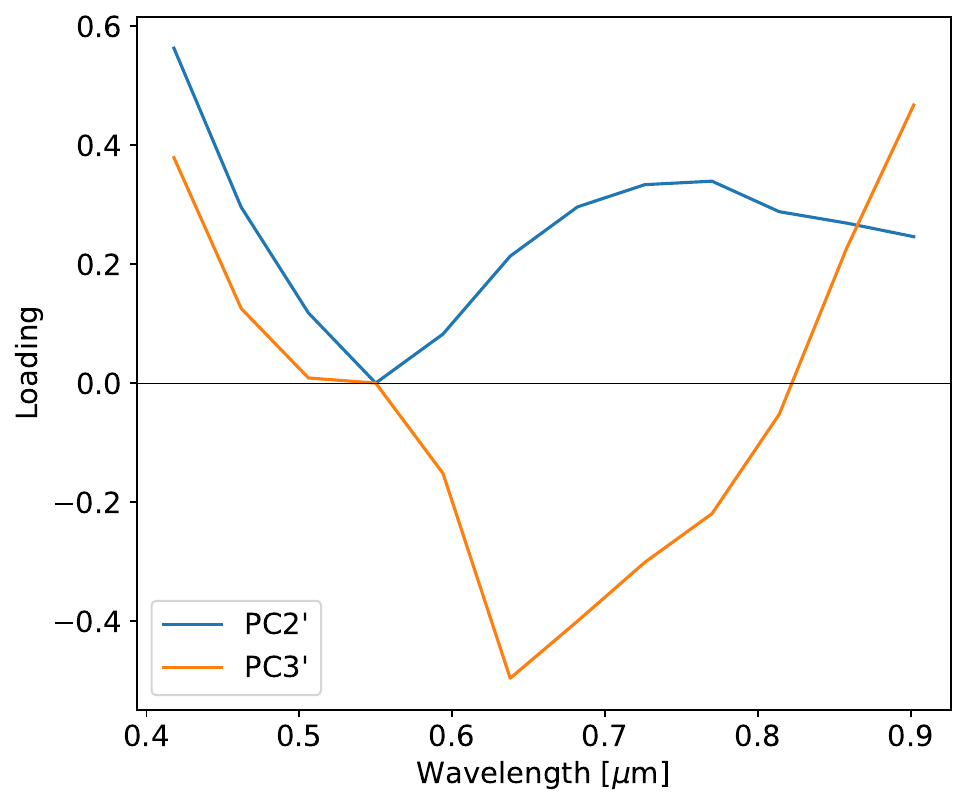}
    \includegraphics[width=0.49\textwidth]{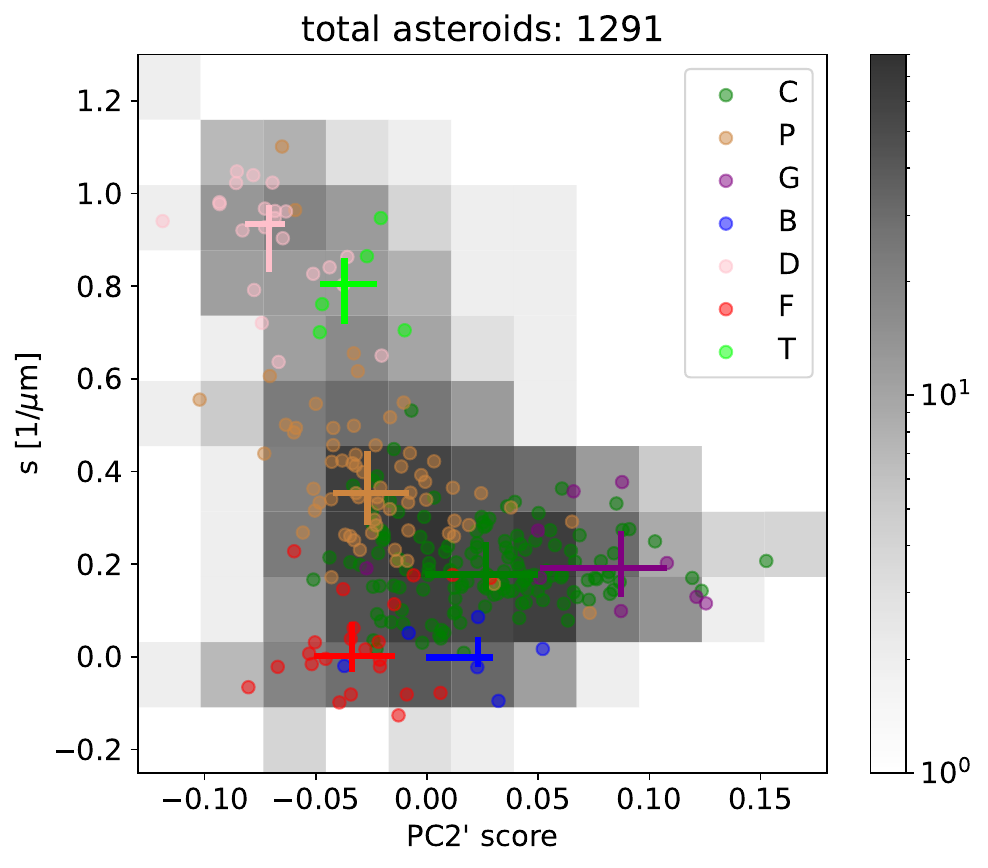}
    \caption{Results from the PCA applied over the reflectance spectra. The upper panel shows the two first components of our PCA. These loading values (elements of the eigenvectors, plotted as a function of wavelength) are used to weight the individual $Gaia$ bands when calculating principal component scores. In the lower panel, we show the principal component space of visible slope $s$ versus PC2' score. The grey hatch represents the density distribution of the 1,290 $Gaia$ asteroids that have good quality data between 0.418 to 0.902 $\mu$m. Those classified by Tholen are overplotted with color markers. Crosses correspond to the median for each taxonomy and the size of the error bars is the interquartile range.}
    \label{fig:PCA}
\end{figure}

After inspecting the general behavior of the different spectral slopes and their combinations, we carried out a PCA over the 12 available reflectance bands between 0.418 and 0.902 $\mu$m similar to what was done by \cite{2002Icar..158..106B}. Although the initial datasets are different, as \cite{2002Icar..158..106B} are including also non-primitive asteroids, their PC3' shape in the principal component base is similar to the shape of our PC2' (ours is shown in the upper panel of Fig. \ref{fig:PCA}, whereas theirs is in Fig. 4 of \cite{2002Icar..158..106B}). We used only dark objects, where the UV absorption and the 0.7 $\mu$m are shaping most of the spectra (mainly G- and C-types), and this is represented by our PC2'. In the same way, they found that "the transformation involving PC3' is especially useful" in isolating  asteroids with spectra of "either a strong UV absorption and shortward of 0.55 $\mu$m or a 0.7-$\mu$m phyllosilicate feature." Our results show that the slope ($s$) and the PC2' are enough to spread the different primitive taxonomies. As PC2' represents the UV and the 0.7 $\mu$m absorptions, in the lower panel of Fig. \ref{fig:PCA}, the G-types and some C-types are those with the largest PC2' values, and the taxonomies having linear spectra in the visible and the smaller absorption up to 0.4 $\mu$m have the lowest PC2' scores.
 
\begin{figure}
    \centering
    \includegraphics[width=0.49\textwidth]{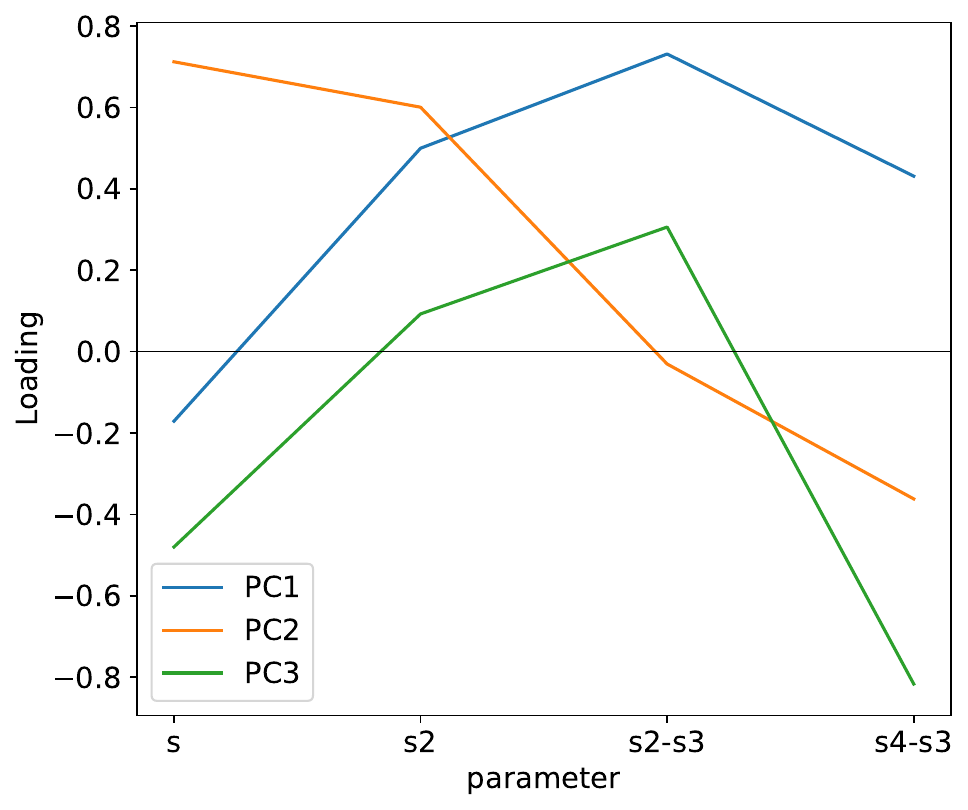}
    \includegraphics[width=0.49\textwidth]{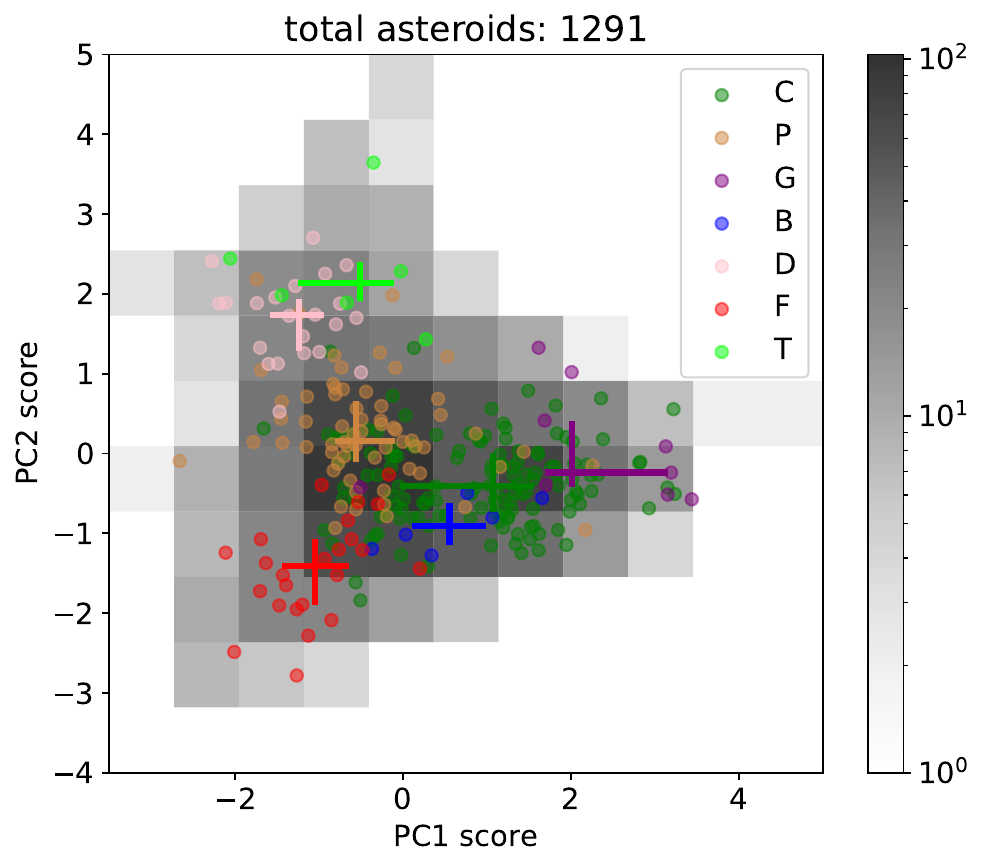}
    \caption{Results from the PCA applied over the spectral parameters: $s$, $s2$, $s2$-$s3$, and $s3$-$s4$. The upper panel shows the first three components of the PCA. These loading values (elements of the eigenvectors) are used to weight each parameter when calculating principal component scores. In the lower panel, we show the principal component space of PC2 versus PC1. The grey hatch represents the density distribution of the 1290 $Gaia$ asteroids that have good quality data between 0.418 and 0.902 $\mu$m. Those classified by Tholen are overplotted with color markers. Crosses correspond to the median for each taxonomy and the size of the error bars is the interquartile range.}
    \label{fig:PCA_ftrs}
\end{figure}

We decided to apply a PCA over selected features, that describe the different types of primitive spectra. Those features are visible slope ($s$) as defined before; $s2$, that allows to easily differentiate F-types from the rest of taxonomies (see Fig. \ref{fig:slope_hist}); $s2-s3$, which is a proxy of the beginning of the UV absorption and presents an interesting bi-modality (see Fig. \ref{fig:slope_hist}); and $s4-s3$, that is a proxy of the 0.7 $\mu$m absorption band. The results of the PCA are shown in Fig. \ref{fig:PCA_ftrs}. In the lower panel, we can see a very similar distribution to the one shown in the lower panel of Fig. \ref{fig:PCA}. This is expected, as PC2 is weighting the slope more than any other feature (see the upper panel of Fig. \ref{fig:PCA_ftrs}) and PC1 is taking care of the NUV and the 0.7 $\mu$m absorptions. However, if we look carefully we can see how the PC2 preserves the bi-modality observed for $s2-s3$ in Fig. \ref{fig:slope-slope_hist}, separating C-types in two regions, one shared with P-types and the other with G-types.

\begin{figure*}
    \centering
    \includegraphics[width=0.39\textwidth]{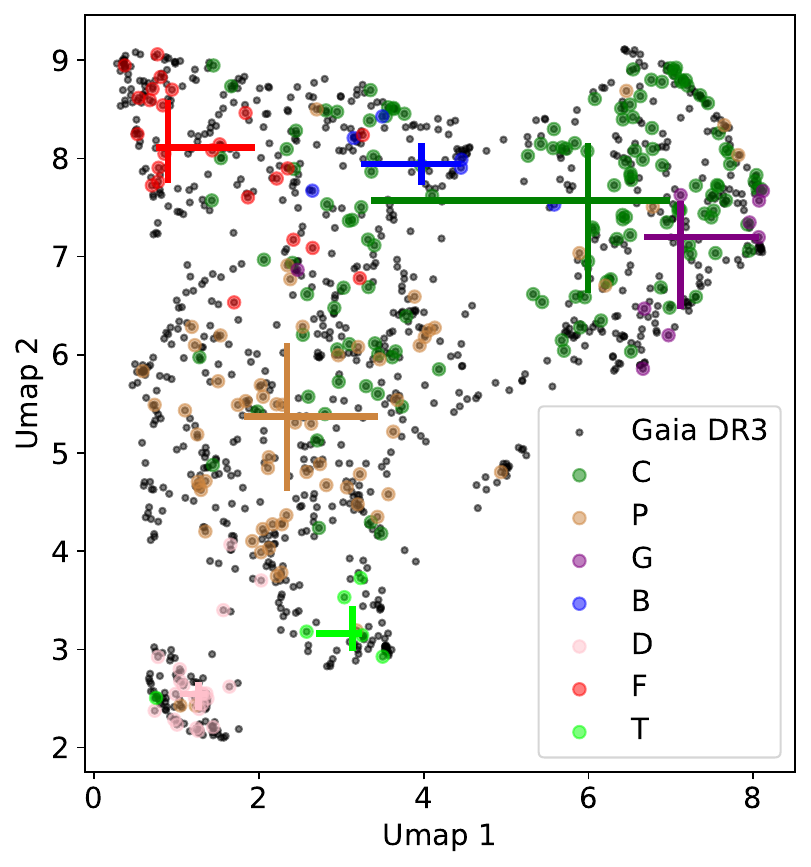} \hspace{0.09\textwidth}
    \includegraphics[width=0.47\textwidth]{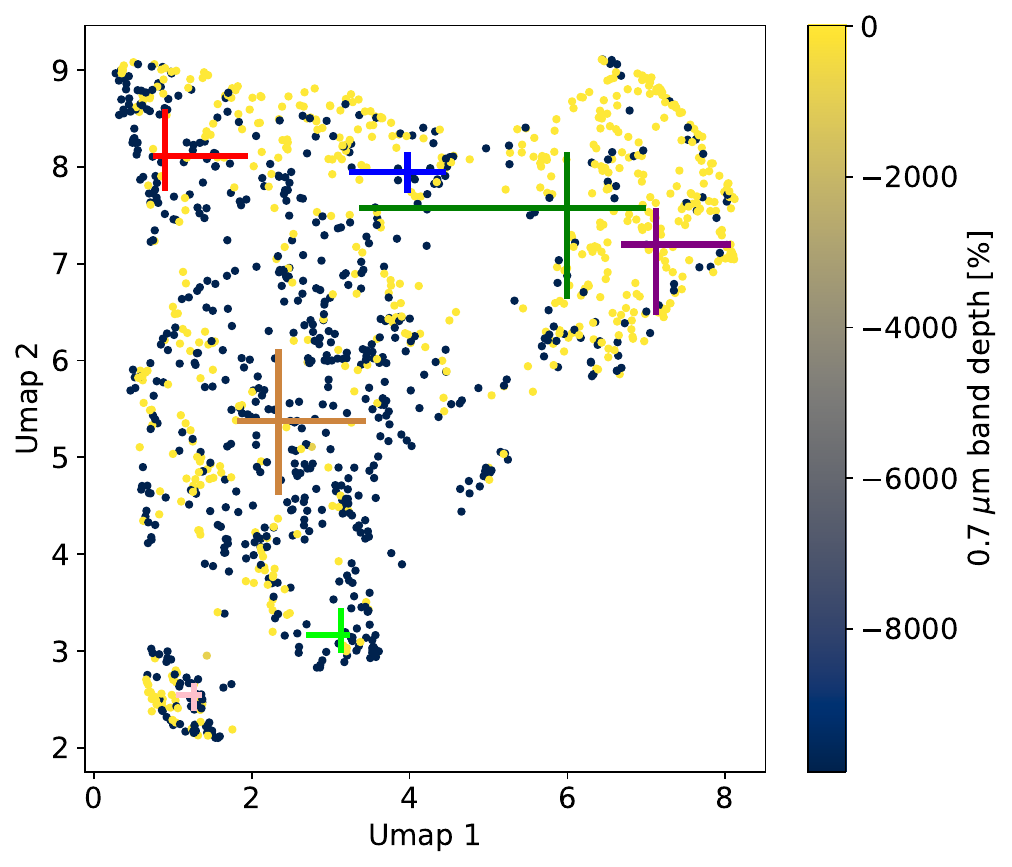} \hspace{0.05\textwidth}\\
    \includegraphics[width=0.49\textwidth]{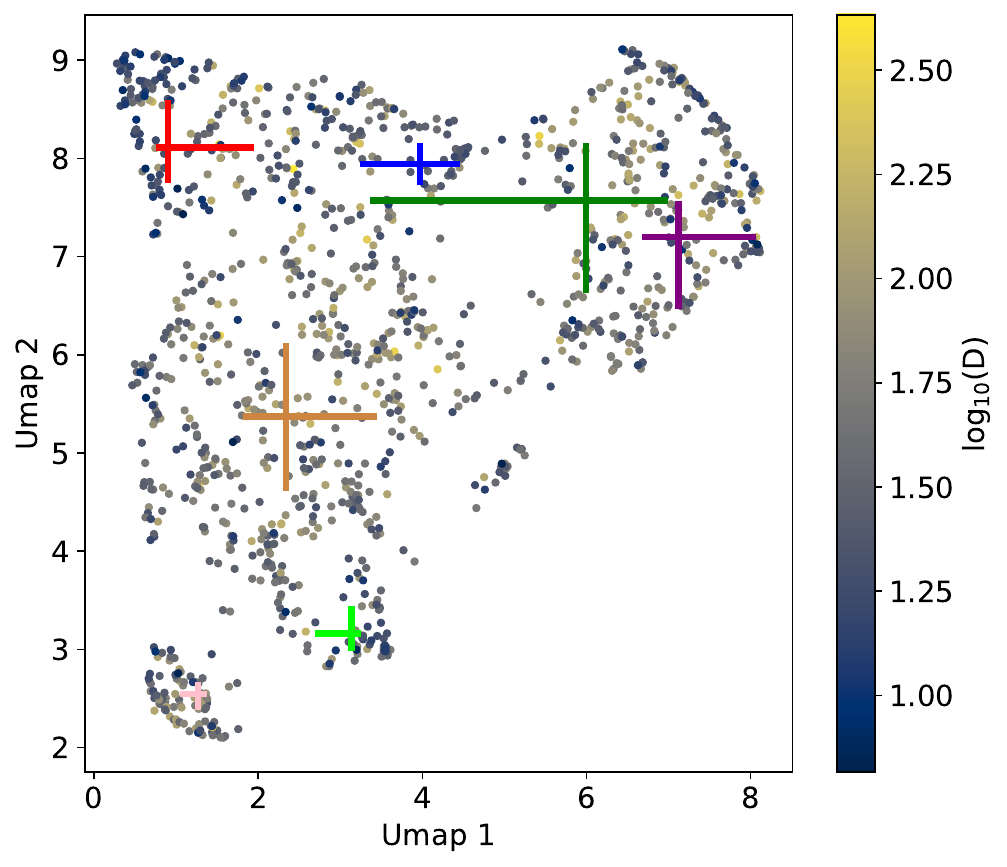}
    \includegraphics[width=0.49\textwidth]{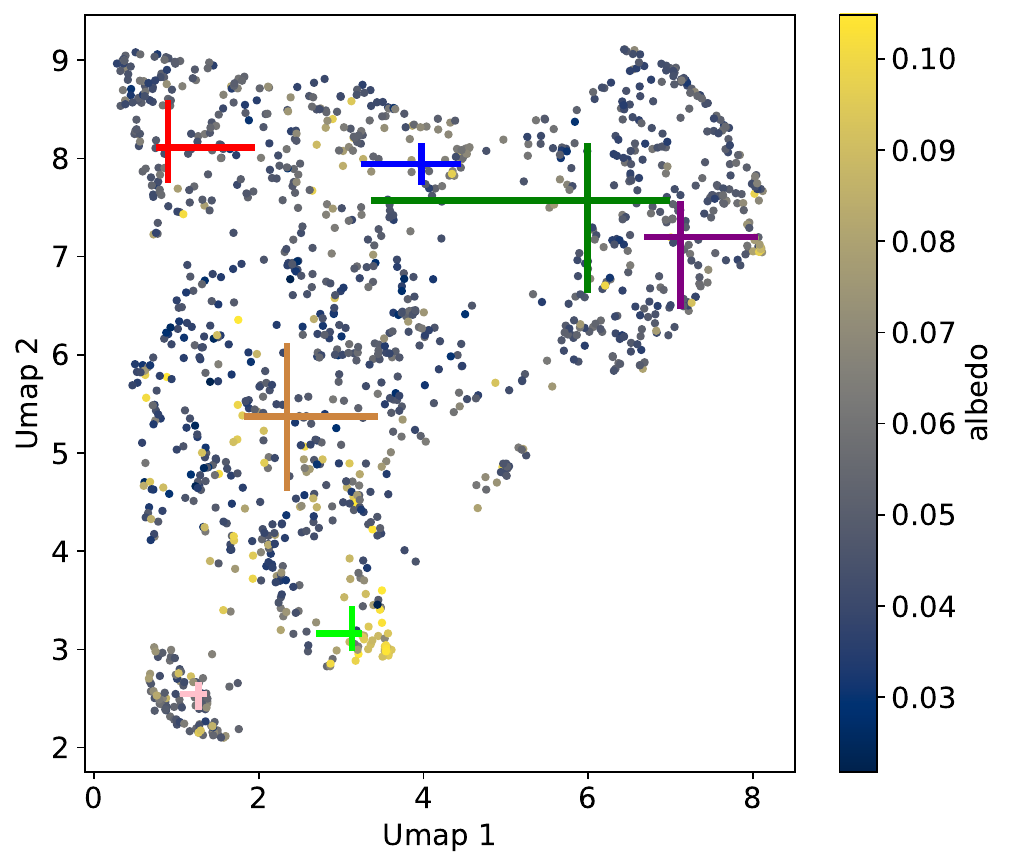} 
    \caption{2D representation of feature space composed by $s$, $s2$, ($s2-s3$), and ($s4-s3$), after applying UMAP. The upper left panel shows all available $Gaia$ spectra in grey, and those classified by Tholen are overplotted in colors following the taxonomy code. Crosses correspond to the median for each taxonomy and the size of the error bars is the interquartile range in all the panels. In the upper-right and lower panels, we have added the following information: the computed depth of the 0.7 $\mu$m band (in \%), the logarithm of the asteroid diameter in kilometers, and the albedo, with an associated color scale shown in the right of each panel. The axes do not have a physical interpretability due to the nonlinearity of the method. From Figure \ref{fig:umap_base}, we can infer the physical properties of the UMAP space based on the parameters used to compute the space.}
    \label{fig:umap}
\end{figure*}

Trying to go further in this analysis, we explored another dimensionality-reduction algorithm, the UMAP. The resulting 2D representation for all the available $Gaia$ spectra (in grey) is shown in the upper left panel of Fig. \ref{fig:umap}, together with those asteroids classified by Tholen. The other panels in the same figure show the values of the computed depth of the 0.7 $\mu$m absorption band (in \%), the logarithm of the asteroid diameter in kilometers, and the asteroid albedo values, displayed with a color scale shown to the right of each panel. It is important to remark that the axes do have not a simple physical meaning due to the nonlinearity of the method. However, we can infer the spectral properties of a region in the UMAP space based on the selected parameters by looking at \ref{fig:umap_base}. Using UMAP, the bimodality observed for the C-types becomes clearer than in any other previous representation. It is important to remark that asteroids that are closer in this 2D space are going to be closer in the original 4D space defined by $s$, $s2$, ($s2-s3$), and ($s4-s3$). We have to note that the shape of the distribution is quite similar to the one obtained with the PCA following a flip in the vertical axis. In this UMAP 2D representation, we can separate most of the taxonomical classes. In particular, C-types seem to be distributed in two clusters: 1) C-types that have an absorption band at 0.7 $\mu$m, associated with iron-rich phyllosilicates and that share the space with G-types in the upper left panel of Fig. \ref{fig:umap}; 2) and C-types that do not have the band and share the space with P- and B-types. This result is also observed in the upper right panel of Fig. \ref{fig:umap}, where we do not see a smooth transition from low to high values of the band depth, but two clear clusters instead (i.e., either the band is present or is not): the 78\% of the C-G cluster members have a 0.7 $\mu$m band detected by our model fitting, while the 66\% of the members C-P-B region have no band detections. Furthermore, the median band depth in the hydrated C-G members is 3.9\%, deeper than the median depth for the hydrated ones in the C-P-B region (2.15\%). We suggest that the C-G cluster is generated because we are including the NUV down to 0.418 $\mu$m; the absorption in the UV is correlated with the absorption in 0.7 $\mu$m (see Fig. \ref{fig:corr}), and we are selecting for the UMAP the features that better describe the behavior of primitive spectra. All that adds extra information that helps separate the hydrated C-types from the rest.

We also have a few detections of the 0.7-$\mu$m absorption band in the regions populated with F-, B-, and D-types. This is not the first time this feature has been detected in taxonomical types different from C-types. \cite{2016A&A...586A.129M} found that 14.3\%, 28.6\%, and 36.4\% of the X-, B-, and T-types, respectively, in the Erigone collisional family showed this feature. Similar values are also observed for primitive taxonomies in the Sulamitis collisional family \citep{2018A&A...610A..25M}. Furthermore, hydration in other types than C was also found in \cite{2022A&A...665A..26M}, using infrared wavelengths.

We have tried this algorithm also over the whole spectral dataset, but the clustering was not as clear, likely due to the increase in variability in certain bands due to their poorly understood systematics errors that are softened by the slope computation.

\subsection{Albedo and diameter}\label{ssec:res_diam}

Looking at the lower right panel of Fig. \ref{fig:umap}, where the color depends on the albedo, we found that the brightest objects among the primitives are in the region populated with T-types. The asteroids in this region have a median value of albedo of 0.087 versus 0.048 as the median of the rest of the sample. The darkest are among the P- and C-types (with and without the 0.7 $\mu$m band), in good agreement with \cite{2011ApJ...741...90M}. Similarly, and regarding the size at the lower left plot, we found that the F-type region is populated with the smallest asteroids among the primitives, with the median size of the asteroids in this region being 22 km, which is below the percentile 25 of the rest of the sample (26 km). This result was also found by \cite{2022A&A...664A.107T}. Finally, if we look carefully at the cluster of asteroids with 0.7 $\mu$m absorption band in the upper right panel of Fig. \ref{fig:umap}, it seems that those with deeper absorptions (in yellow) have smaller diameters (dark blue in the lower-left panel of Fig. \ref{fig:umap}). 

\begin{figure*}
    \centering
    \includegraphics[width=0.4\textwidth]{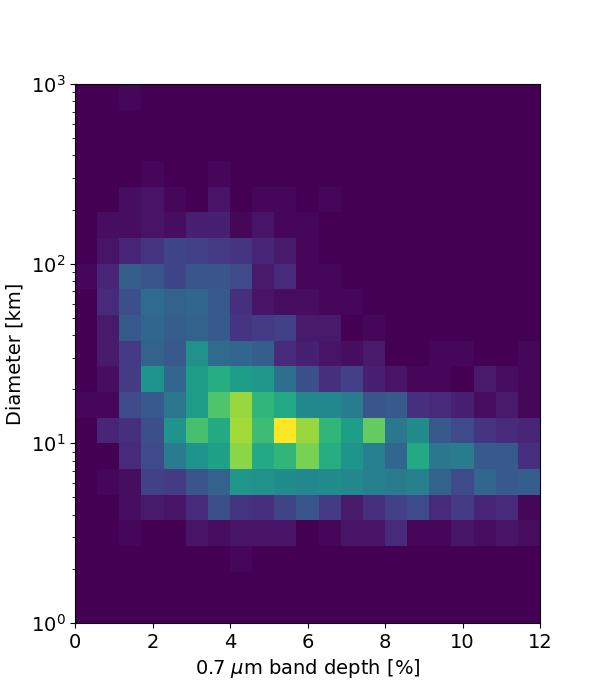}
    \includegraphics[width=0.4\textwidth]{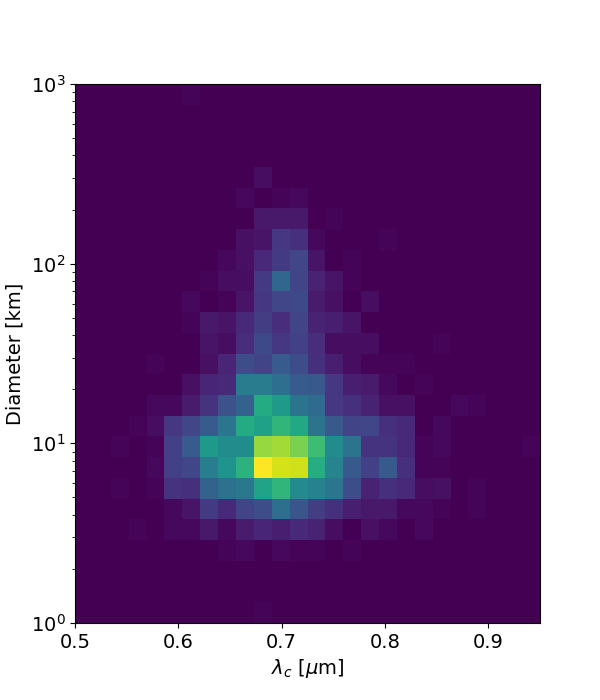}
    \caption{Density plot of the diameter of hydrated asteroids versus their 0.7 $\mu$m band depth (left panel) and the center of their band (right panel).}
    \label{fig:diam_depth}
\end{figure*}

We explore further this apparent trend in the density plot of the diameter versus the depth of the 0.7 $\mu$m absorption band in Fig. \ref{fig:diam_depth}. Although a visual inspection suggests the presence of such a trend, it is important to note that smaller asteroids have lower S/N$_{ref}$ (as defined in Section \ref{sec:sam}) since the $Gaia$ bands (or points) have larger errors for these asteroids. This translates into a larger fraction of unrealistic values for the depth, both negative values and large ones. We know from other studies that the maximum depths observed for the 0.7 $\mu$m band in primitive asteroids are around 7-8\% \citep{2014Icar..233..163F,2016A&A...586A.129M,2018A&A...610A..25M,2019A&A...630A.141M}, so measurements above these values are most likely due to the noise. In the case of the band center, we also see a larger dispersion of values for the smaller sizes in the right panel of Fig. \ref{fig:diam_depth}. Nevertheless, and as previously pointed out by \cite{2014Icar..233..163F}, the visible spectra of hydrated asteroids show other absorption bands, centered at 0.6 and 0.8-0.9 $\mu$m, and associated to charge transfer transitions in iron oxides \citep{1979EcGeo..74.1613H,1985Icar...63..183F,1994Icar..111..456V}, which can contribute to the observed dispersion in the values for the band center. From the results shown above we can only conclude that the average band depths (3.5 $\pm$ 1.0 \%) and band centers (0.70 $\pm$ 0.03 $\mu$m) obtained for the primitive asteroids observed with $Gaia$ are in good agreement with the values found in the literature.

\begin{figure*}
    \centering
    \includegraphics[width=0.99\textwidth]{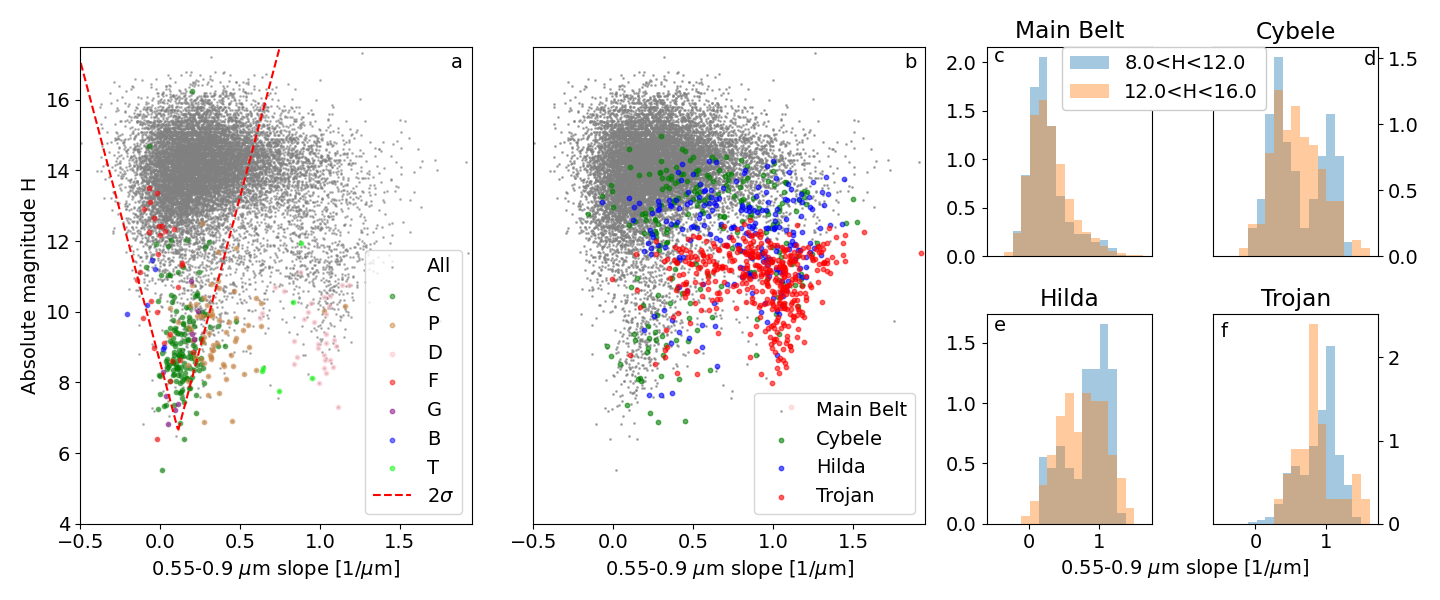}
    \caption{(a) Replication of Figure 7 from \cite{2019A&A...630A.141M}, but using primitive asteroids among the $Gaia$ DR3 dataset. Colors indicate objects classified with Tholen taxonomy. Red lines are the 2$\sigma$ dispersion of the sample as in \cite{2019A&A...630A.141M}, corresponding to their Equation 5. (b) Same as (a) but with main belt asteroids (in grey) separated from Cybeles (in green), Hildas (in blue), and Trojans (in red). See the main text for the semi-major axis values used to separate these populations. (c), (d), (e), and (f) show the visible spectral slope distribution for the four dynamical populations observed for absolute magnitudes $8 < H < 12$ (in blue) and $12 < H < 16$ (in beige).}
    \label{fig:diam_slope}
\end{figure*}

After analyzing the UMAP 2D-distributions we searched for any further relationship between the physical parameters and the slopes. When plotting the visible slope, $s,$ against the asteroid diameter (or asteroid absolute magnitude $H$), we observed what seemed to be an increase in the slope dispersion with decreasing size (increasing $H$), generating what appeared to be two inverted-triangle-shaped distributions (Fig. \ref{fig:diam_slope}a): one centered at the median $s$ of C-types and the other close to the median slope of D-types. The one for C-types is similar to the distribution obtained by \cite{2019A&A...630A.141M} using a smaller sample of primitive asteroids, all of them being members of collisional families of the inner main belt. We note that to properly compare their results with this study, we re-computed the visible spectral slope using the same wavelength range as in the study cited above, namely, from 0.55 to 0.90 $\mu$m. To explain this behavior, they hypothesize that larger asteroids have more heterogeneous surfaces with different grain sizes, crater ages, and compositional units, thus implying different spectral slopes among their surfaces. Their unresolved spectra will result in neutralized spectra with a mean slope of all the material. Meanwhile, smaller asteroids have a more homogeneous surface as they are sampling a smaller fraction of the parent body, thus exhibiting a wider variety of spectral slopes. Furthermore, space weathering is seen to redden some carbonaceous materials, like CV/CO carbonaceous chondrites, while blueing others, as is the case for some types of CI/CM chondrites \citep{2018Icar..302...10L}, contributing to the above-mentioned variety.\\

We included in Fig. \ref{fig:diam_slope}a the 2$\sigma$ region obtained by \cite{2019A&A...630A.141M}, using their Eq. (5), and represented them as two red dashed lines, as well as those asteroids classified using Tholen taxonomy. The two lines define a region and if the observed tendency is real, it is expected that "the spread in spectral slopes is contained within the limits of this region", as explained in \cite{2019A&A...630A.141M}. For the whole sample of primitive asteroids among $Gaia$, we see that 99.1\% of the asteroids lie on the right side of the left straight line, but a large number of objects are also located to the right of the right straight line. Looking at the different taxonomies, this distribution seems to be related to taxonomies such as P-, T-, and in particular, D-types. It seems that D-types (light pink) follow a similar inverted triangular shape to that of the C-types (green) and that there are very few big objects in between the two inverted triangles. To properly interpret this distribution of the visible slope, we first separated the main belt asteroids from the dynamical populations \textbf{beyond the main belt}, namely, the Cybeles (with semi-major axis $3.3 < a < 3.7$ au), the Hildas ($3.7 < a < 4.2$ au), and what we call here the Trojans, which include all asteroids having $a > 4.2$ au. The four populations are plotted with different colors in Fig. \ref{fig:diam_slope}b. It is clear from this plot that the \textbf{population beyond the main belt} is responsible for the second inverted triangle, associated with D-types.

Once we have separated main-belt asteroids from the populations beyond, we have to consider the main belt's actual size-frequency distribution (SFD). It is a well-known fact that there are several "bumps" in the main belt's SFD \citep[see, e.g., Fig. 1 in][]{2005Icar..175..111B} different parts of the distribution \textbf{having different gradients. In particular, a significant change in the gradient of the SFD occurs at }$H$ = 12 ($D \sim$ 20 km for $p_V$ =0.07), which is the value where we start to see many more objects in the belt; \textbf{the cloud of objects} starts to disappear for $H >$ 14-15, where we start to have an actual observational bias (problems to detect small asteroids). In the region between 20 and 100 km (i.e., 8 $< H <$ 12), the \textbf{gradient} of the SFD is significantly \textbf{smaller} (almost flat) and we have a density of asteroids that increases slowly with $H$. Indeed, if we compute the distribution of the visible spectral slope for 8 $< H <$ 12 and 12 $< H <$ 16 (shown in Fig. \ref{fig:diam_slope}c), we can see that the two distributions overlap. The median value of the distribution for larger asteroids is $0.21^{+0.29}_{-0.18}$ $\mu$m, meanwhile for the distribution for the smaller asteroids is $0.25^{+0.32}_{-0.22}$ $\mu$m. We, therefore, conclude that the trend between size and visible slope observed by \cite{2019A&A...630A.141M} for the inner belt primitive families is not observed for the entire main belt population; instead, the dispersion of spectral slope values observed for the smaller asteroids is simply due to the actual SFD of the main belt. A further analysis of the primitive families will be done with the DR4.

We performed the same exercise for the dynamical populations \textbf{beyond the main belt}, with their corresponding histograms (also shown in Fig. \ref{fig:diam_slope}d, e, and f). For the Hildas and Cybeles, we observed a bimodal distribution for 8 $< H <$ 12, with the objects concentrating at spectral slopes between 0.2 and 0.6 $\mu$m$^{-1}$ (P-types, on average) and spectral slopes $>$ 0.6 $\mu$m$^{-1}$ (T-types and D-types). This behavior has been previously observed by other authors \citep{2008Icar..193...20S,2010Icar..206..729G,2018Icar..311...35D}. For 12 $< H <$ 16, they are equally distributed among D- and P- taxonomies. To explain the larger fraction of P-types at $H < $12 observed for the Hildas, these authors invoked a combination of space weathering and resurfacing, and a wavy size distribution produced by the Hildas being in a stable 3:2 resonance with Jupiter, surrounded by unstable regions. This process, however, does not work for the Cybeles, since these objects are not orbiting in a mean motion resonance. 

In the case of the Trojans, \textbf{ we obtained the well-known bimodal spectral slope distribution for asteroids with 8 $< H <$ 12, which was not observed for asteroids with 12 $< H <$ 16 (Fig. \ref{fig:diam_slope}f). This behavior was also detected by \cite{2015AJ....150..174W} for the L4 swarm, using data from SDSS and an 8.2m Subaru telescope. In addition, the two slope distributions are slightly different, with the bigger asteroids (8 $< H <$ 12) having slightly redder slopes than the smaller ones (12 $< H <$ 16). This was also observed by \cite{2008A&A...483..911R} for the background Trojan asteroids (i.e., not members of a family), also having high orbital inclinations ($\sin{i} >$ 0.2) and by \cite{2015AJ....150..174W} for the L4 swarm. In both cases, the authors invoked collisional evolution, either by fragmentation or by collisional resurfacing (red objects converted to less-red objects upon collision) to explain their findings. They concluded that collision may have not played a major role in shaping the magnitude distribution of the bigger Trojans since captured by Jupiter.}

\subsection{Families}\label{ssec:res_fam}

\begin{figure}
    \centering
    \includegraphics[width=0.49\textwidth]{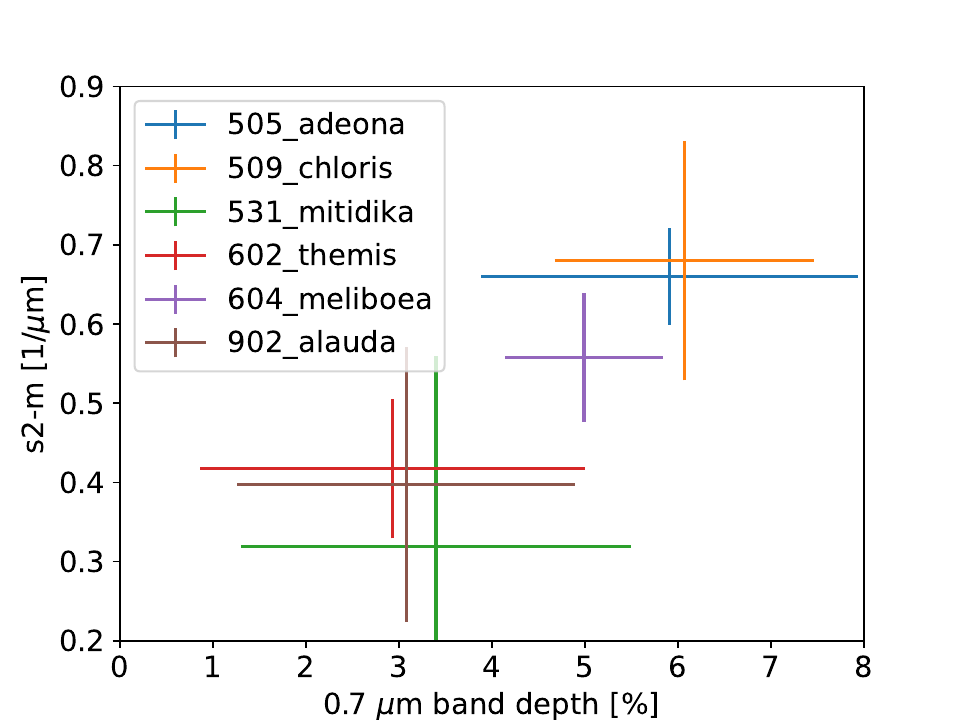}
    \caption{Scatter plot of the median values of $s2-m$ ($s2$ described in Fig. \ref{fig:ref_sp} and $m$ in Eq. \ref{eq:fitting_model}) and the 0.7 $\mu$m band depth for the primitive families that have 5 or more asteroids with computed $s2$ and band depth bigger than zero (Adeona: 12, Chloris: 5, Mitidika: 5, Themis: 35, Meliboea: 11, and Alauda 15). The size of the crosses is the dispersion of the values of the families.}
    \label{fig:fam}
\end{figure}

Collisional families are believed to be the direct outcome of a disruptive and/or cratering event and thus, their members are expected to be composed of similar material and to be exposed at a similar time to space. In Fig. \ref{fig:fam}, we present with colored crosses the median values of $s2-m$ (with $m$ as the slope obtained from our fitting model described in Eq. \ref{eq:fitting_model}) and the depth of the 0.7$\mu$m absorption band for the members of a total of six collisional families, all from the $Gaia$ sample. The size of the crosses corresponds to the dispersion of the values. We show here only those families having five or more members with computed $s2$ slope and with a 0.7$\mu$m band depth bigger than zero: (Adeona: 12, Chloris: 5, Mitidika: 5, Themis: 35, Meliboea: 11, and Alauda: 15). The slope difference $s2-m$ is indicative of the strength of the UV absorption. We observe, within the dispersion of the points, a correlation between that strength and the depth of the 0.7 $\mu$m absorption band, with the families having a stronger UV absorption presenting a deeper band. This result reinforces the findings described in Section \ref{ssec:res_gen}. Again, a further analysis of the primitive families will be carried out with DR4.

\section{Conclusions}\label{sec:con}

In this work, we have aimed to make a detailed study of the spectral features of primitive asteroids using the $Gaia$ DR3, paying special attention to the UV absorption feature, which can be inferred below 0.4 -- 0.5 $\mu$m. We analyzed 15,529 asteroids over the spectral range from 0.374 to 0.902 $\mu$m. Our main results can be summarized as follows:
\begin{itemize}
    \item For C, P, and G-type asteroids, the 0.7-$\mu$m absorption band is correlated with the UV absorption when including the bluest part of the spectra. These two absorption features are signatures of iron-rich phyllosilicates and indicative of high degrees of hydration in the minerals.\\
    \item Using combinations of slopes as descriptors of the reflectance spectra in both PCA and UMAP (dimensionality-reduction algorithms), we observed a bimodality distribution of the C-types. One cluster displayed the 0.7 $\mu$m band and the UV absorption, as with the G-types, and the other one did not display the band at all. \\
    \item F-types, which could not be identified without the use of the NUV wavelengths, display the beginning of the UV absorption around 0.4 $\mu$m. We also found hints that they are consistently smaller than other taxonomies. To confirm these results, we would need further exploration in the laboratory as well as more observations containing the UV-blue part of the spectra.\\
    \item We obtained average values for band depth (3.5 $\pm$ 1.0 \%) and central wavelength (0.70 $\pm$ 0.03 $\mu$m) for the hydrated objects in $Gaia$ DR3. This is in good agreement with previous works. We noticed that for smaller asteroids the reflectance spectra are noisier and the obtained values of depth and central wavelength have a larger dispersion.\\
    \item We reproduced the $H$ vs. visible slope presented in \cite{2019A&A...630A.141M} for the primitive families of the inner asteroid belt. They claimed larger dispersion of slope values for smaller asteroids is not observed when using the entire population of primitive main belt asteroids among $Gaia$ DR3. For the populations beyond the main belt, namely, Cybeles, Hildas, and Trojans, we found similar slope distributions as those found the literature, with both Cybeles and Hildas presenting bimodal distributions for $H < 12$.
\end{itemize}

All these results shed light on the importance of having spectral data at the shortest wavelengths to better characterize the primitive taxonomies, showing that this is a promising region for diagnosing the composition of the primitive asteroids.\\

\begin{acknowledgements}
FTR, JdL, and JL acknowledge support from the Agencia Estatal de Investigaci\'{o}n del Ministerio de Ciencia e Innovaci\'{o}n (AEI-MCINN) under the grant "Hydrated Minerals and Organic Compounds in Primitive Asteroids" with reference PID2020-120464GB-100.\\
FTR also acknowledges the support from the COST Action that fully found his stay at OCA where he began to study the Gaia dataset. He also acknowledges the OCA group of asteroids for their support and welcome during those days. He also acknowledges the big effort by JdL and ET in this work.\\
\end{acknowledgements}

\bibliographystyle{aa}
\bibliography{main}

%
%
\onecolumn 
\begin{appendix}
\section{References of visual albedos and diameters}\label{sec:app_ref}
\begin{table*}[h]
    \caption{Data references used for albedos and diameters.}
    \centering
    \begin{tabular}{p{18cm}}
    \\
    \hline
\text{\cite{2013A&A...554A..71A,2016A&A...591A..14A,2018A&A...612A..85A,2020A&A...638A..84A}}; \text{\cite{2017A&A...603A..55A}}; \text{\cite{2014MNRAS.443.1802B}}; \text{\cite{2013ApJ...773...22B}}; \text{\cite{2014Icar..239..118B,2023A&A...671A.151B}}; \text{\cite{2021MNRAS.502.4981C}}; \text{\cite{1999Icar..140...53C}}; \text{\cite{2004PhDT.......371D}}; \text{\cite{2009P&SS...57..259D}}; \text{\cite{2014A&A...564A..92D}}; \text{\cite{2011Icar..214..652D}}; \text{\cite{2013A&A...555A..15F}}; \text{\cite{2011ApJ...742...40G, 2012ApJ...744..197G, 2012ApJ...759...49G}}; \text{\cite{2013PASJ...65...34H}}; \text{\cite{2013Icar..226.1045H,2015Icar..256..101H,2017A&A...599A..36H,2017A&A...601A.114H,2018Icar..309..297H,2020A&A...633A..65H}}; \text{\cite{2019pdss.data....3H}}; \text{\cite{2012MNRAS.423.2587H}}; \text{\cite{2022PSJ.....3...56H}}; \text{\cite{2021AJ....162...40J}}; \text{\cite{2016ApJ...817L..22L}}; \text{\cite{2016A&A...585A...9L}}; \text{\cite{2007Icar..186..152M,2007Icar..186..126M}}; \text{\cite{2011ApJ...743..156M, 2012ApJ...760L..12M}}; \text{\cite{2006Icar..185...39M,2012Icar..221.1130M}}; \text{\cite{2023A&A...670A..52M}}; \text{\cite{2011ApJ...741...68M,2012ApJ...759L...8M,2014ApJ...791..121M,2017AJ....154..168M,2020PSJ.....1....5M,2021PSJ.....2..162M}}; \text{\cite{2021A&A...652A.141M}}; \text{\cite{2022PSJ.....3...30M}}; \text{\cite{2015ApJ...814..117N,2016AJ....152...63N}}; \text{\cite{2012Icar..221..365P}}; \text{\cite{2016Sci...353.1008R}}; \text{\cite{2015A&A...578A..42R}}; \text{\cite{2010AJ....140..933R,2011AJ....141..186R}}; \text{\cite{2008Icar..193...20S}}; \text{\cite{2019Sci...364..252S}}; \text{\cite{2002AJ....123.1056T}}; \text{\cite{1999Icar..140...17T}};  \text{\cite{2010AJ....140..770T}}; \text{\cite{2011PASJ...63.1117U}}; \text{\cite{2020NatAs...4..136V,2021A&A...654A..56V}}; \text{\cite{2017A&A...607A.117V}};  \text{\cite{2019Sci...364..268W}};\text{\cite{2020A&A...641A..80Y}}\\
    \hline
    \end{tabular}
    \label{tab:alb_ref}
\end{table*}
    
\section{UMAP results}
\begin{figure*}[h]
    \centering
    \includegraphics[width=0.4\textwidth]{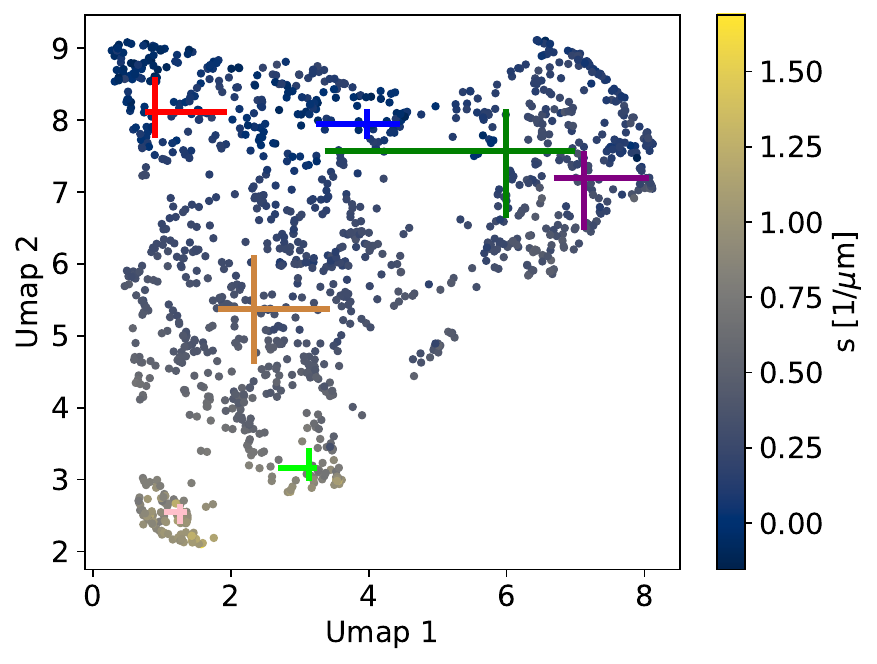}
    \includegraphics[width=0.4\textwidth]{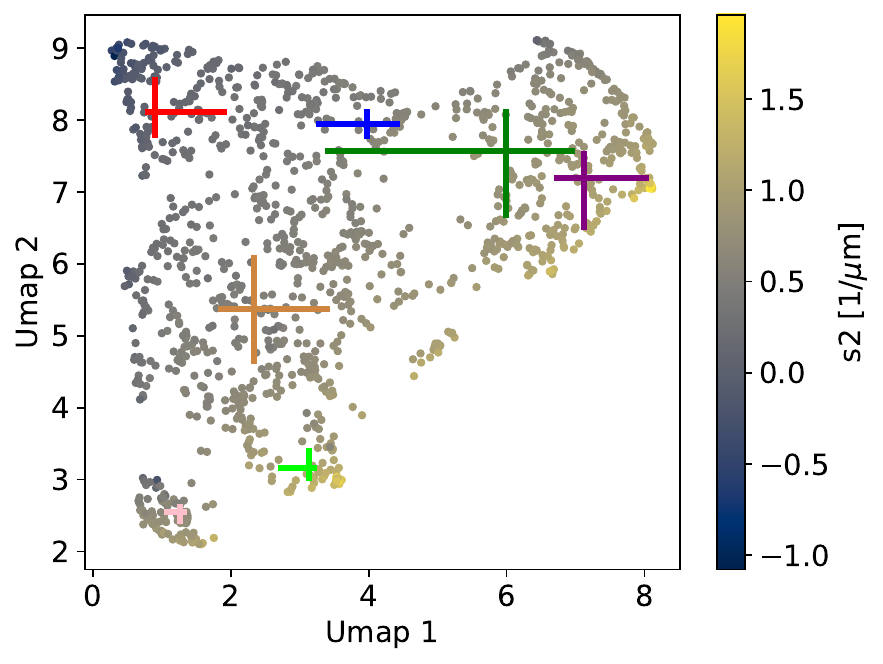}
    \includegraphics[width=0.4\textwidth]{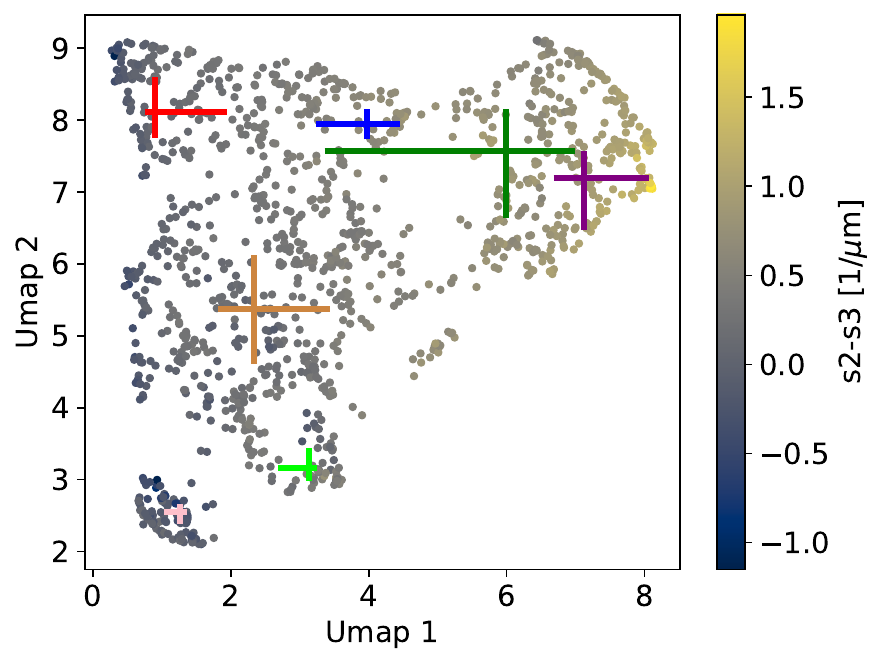}
    \includegraphics[width=0.4\textwidth]{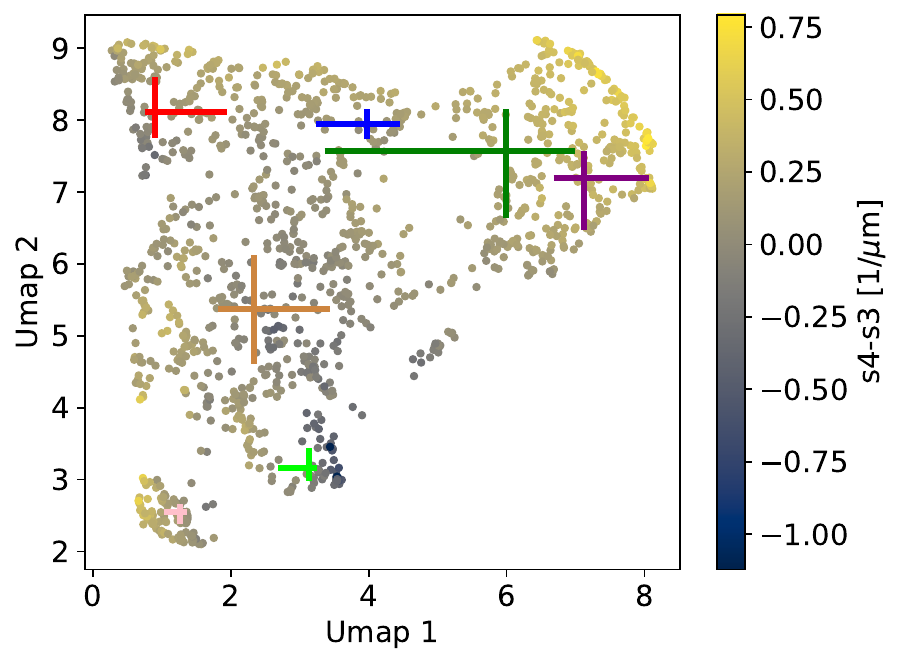}
    \caption{Same as Fig. \ref{fig:umap} but using the four spectral features described in the text: $s$, $s2$, $s2$-$s3$, and $s4$-$s3$. Crosses correspond to the median for each taxonomy (following the same color code) and the size of the error bars is the interquartile range.}
    \label{fig:umap_base}
\end{figure*}

In Fig. \ref{fig:umap_base}, we show the resulting UMAP 2-D representation, but adding the following information, from top to bottom, left to right: $s$, $s2$, ($s2$-$s3$), and ($s4$-$s3$). A color scale is shown to the right of each panel, as in Fig. \ref{fig:umap}.

\end{appendix}

\end{document}